\documentclass[useAMS,usenatbib]{mn2e}
\pdfoutput=1
\usepackage[ansinew]{inputenc}
\usepackage{amsfonts}
\usepackage{amsmath}
\usepackage{enumerate}
\usepackage{enumitem}
\usepackage{graphicx}
\usepackage{epsfig}
\usepackage{color}
\usepackage{amssymb}

\newcommand{\eg}{{e.g., }}
\newcommand{\ie}{{i.e., }}
\newcommand{\pz}{photo-$z$\ }
\newcommand{\pzns}{photo-$z$} 
\newcommand{\PZ}{Photo-$z$\ }

\title[TPZ : Photometric redshift PDFs by using prediction trees and random forests]
{TPZ : Photometric redshift PDFs and ancillary information by using prediction trees and random forests}
\author[M. Carrasco Kind and R. J. Brunner] 
{Matias Carrasco Kind\thanks{E-mail: mcarras2@illinois.edu} and Robert J. Brunner\\
Department of  Astronomy, University of Illinois, Urbana, IL 61820 USA}

\begin{document}
\date{\today}

\pagerange{\pageref{firstpage}--\pageref{lastpage}} \pubyear{0000}

\maketitle

\label{firstpage}
\begin{abstract}
With the growth of large photometric surveys, accurately estimating photometric redshifts, preferably as a probability density function (PDF), and fully understanding the implicit systematic uncertainties in this process has become increasingly important. In this paper, we present a new, publicly available, parallel, machine learning algorithm that generates photometric redshift PDFs by using prediction trees and random forest techniques, which we have named TPZ. This new algorithm incorporates measurement errors into the calculation while also dealing efficiently with missing values in the data. In addition, our implementation of this algorithm provides supplementary information regarding the data being analyzed, including unbiased estimates of the accuracy of the technique without resorting to a validation data set, identification of poor photometric redshift areas within the parameter space occupied by the spectroscopic training data, a quantification of the relative importance of the variables used 
to construct the PDF, and a robust identification of outliers. This extra information can be used to optimally target new spectroscopic observations and to improve the overall efficacy of the redshift estimation. We have tested TPZ on galaxy samples drawn from the SDSS main galaxy sample and from the DEEP2 survey, obtaining excellent results in each case. We also have tested our implementation by participating in the PHAT1 project, which is a blind photometric redshift contest, finding that TPZ performs comparable to if not better than other empirical photometric redshift algorithms. Finally, we discuss the various parameters that control the operation of TPZ, the specific limitations of this approach and an application of photometric redshift PDFs.
\end{abstract}


\section{Introduction} 
Late time cosmological measurements are often made by carefully measuring the three-dimensional distribution of galaxies. For these measurements, the distance between the galaxy and the observer is most accurately made by using a spectroscopic redshift. However, spectroscopic measurements are considerably more difficult to obtain, and are, therefore, more expensive than photometric measurements, as they require long exposures in order to achieve sufficient signal-to-noise over a wide wavelength range. As an example, while the Sloan Digital Sky Survey (SDSS; \citealt{York2000}) has taken millions of spectroscopic redshifts of galaxies to high precision~\citep{Aihara2011}, the same survey has obtained detailed photometric measurements for a much larger sample of galaxies in considerably less time. This dichotomy will only grow with ongoing and planned surveys that are dominated by photometric-only observations.

As a result, considerable attention has been focused on the estimation of redshifts by applying statistical techniques to the photometric observations of sources through different filters. These photometric redshift (hereafter \pzns) estimation techniques have become crucial for modern, multi-band digital
 surveys; and this need for fast and accurate \pz estimation is becoming even more important for large photometric surveys like the Dark Energy Survey (DES\footnote{http://www.darkenergysurvey.org/}) and the Large Synoptic Survey Telescope (LSST\footnote{http://www.lsst.org/lsst/}), which are probing galaxies that are often too faint to be spectroscopically observed. Adopting a \pz approach allows cosmological measurements on galaxy samples that are currently at least a hundred times larger than comparable spectroscopic samples, that have relatively simple and uniform selection functions, and that extend to fainter flux limits and larger angular scales and thus probe much larger cosmic volumes. In summary, \pz techniques provide a much higher number of galaxies with redshift estimates per unit telescope time than spectroscopic surveys~\citep{Hildebrandt2010}.

The estimation of galaxy redshifts using multi band photometry was first performed by~\cite{Baum1962}, while~\cite{Koo1985} and~\cite{Loh1986} were the first to compute galaxy redshifts by using digital photometric observations from charge coupled devices. In the last fifteen years, however, the estimation of redshifts from broadband photometry has grown significantly. Presently, there are many different methods for computing photometric redshifts~\citep[see, \eg][for an updated comparison of current photometric redshift methods and public codes]{Wang2008,Hildebrandt2010,Abdalla2011}. These techniques can be broadly categorized as either template fitting algorithms or empirical training algorithms. The template fitting algorithms~\citep[\eg][]{Benitez2000,Bolzonella2000,Csabai2003,Ilbert2006,Feldmann2006,Assef2010} can either use empirical~\citep[\eg][]{Coleman1980,Assef2010} or synthetic spectral templates~\citep[\eg][]{Bruzual2003}. These techniques estimate a photometric redshift by finding the best match 
between the 
observed 
magnitudes or colors and the synthetic magnitude or colors from the suite of templates that are sampled across the expected redshift range of the photometric observations.
 
Empirical training methods use a spectroscopic training data set to calibrate an algorithm that can be quickly applied to new photometric observations. Initially the training set was used to map a polynomial function between the colors and the redshift~\citep[\eg][]{Connolly1995,Brunner1997}. More recently, this process has been extended to machine learning algorithms, including artificial neural networks~\citep[\eg][]{Collister2004,Oyaizu2008a}, boosted decision trees~\citep[\eg][]{Gerdes2010}, random forest~\citep[\eg][]{Carliles2010}, nearest neighbors~\citep[\eg][]{Ball2007,Ball2008,Lima2008}, self-organized maps~\citep[\eg][]{Geach2012,Way2012}, spectral connectivity analysis~\citep[\eg][]{Freeman2009}, Gaussian process~\citep[\eg][]{Way2009,Bonfield2010}, support vector machines~\citep[\eg][]{Wadadekar2005} or Quasi Newton Algorithm~\citep[\eg][]{Cavuoti2012}. While only a few of these \pz methods are publicly available, they all perform to a similar accuracy and provide only a single redshift estimate 
rather than a full redshift probability density function for each galaxy.

The template fitting methods, which leverage model galaxy spectral energy distributions (SED), have been used extensively and are often preferred since once implemented they can be readily applied to new data by simply adopting the appropriate photometric filter transmission functions. Given a representative sample of template galaxy spectra, most of these techniques can reliably predict a \pzns, although the use of training data that includes known redshifts can improve these predictions~\cite[\eg][]{Benitez2000,Ilbert2006}. These techniques, however, are not exempt from uncertainties due to measurement errors on the survey filter transmission curves, mismatches when fitting the observed magnitudes or colors to template SEDs, and color-redshift degeneracies. Furthermore, template techniques generally become less reliable at high redshift where  the uncertainties in galaxy SEDs increases, since the templates are often calibrated using low redshift galaxies. 

On the other hand, when provided with a high quality spectroscopic training sample, empirical training techniques have been shown to have similar or even better performance~\citep{Collister2004}. In addition, empirical techniques are generally simpler to apply to different data sets and frequently provide an improved quantification of any uncertainties, which can be encoded in a \pz probability  density function (PDF). They also have the additional advantage that is easier to include extra information, such as galaxy profiles, concentration, angular sizes, or environmental properties, in addition to magnitudes or colors. These methods, however, are only reliable within the limits of the training data, and sufficient caution must be exercised when extrapolating these algorithms beyond the limits of the training data.

As the demand for more accurate \pz methods has grown, techniques have branched out into new areas in order to improve the accuracy of \pz estimation. While a complete understanding of the systematic uncertainties is needed for a reliable and accurate machine learning photo-$z$ algorithm~\citep[see, \eg][for a discussion on photometric redshift errors]{Oyaizu2008b}, other issues have recently been recognized in the effort to generate the most accurate photometric redshifts. For example,~\cite{Cunha2012a,Cunha2012b} analyzed the effect of systematics within the spectroscopic training data set that is used to estimate a galaxy \pzns. Likewise, other functionality that a modern \pz algorithm should provide include an identification of outliers on the training set that lead to an incorrect estimation of a \pzns, an identification of the features within the training data that most strongly affect a \pz estimate, and an identification of areas of parameter space (\eg magnitudes, colors, and redshift ranges) that 
are 
under sampled by the training data. The last two features are important to the design of photometric surveys, as they provide useful information to optimally and efficiently guide follow-up spectroscopy to generate the scientifically most useful training data set for these algorithms.

Of course, we estimate a galaxy's redshift so that it can be used in a subsequent analysis.  A number of cosmological measurements such as galaxy clustering, weak lensing, baryon acoustic oscillations and the mass function of galaxy clusters~\citep[see, \eg][]{Ho2012,Reid2010,Jee2013}, among others, depend strongly on both the number of targeted galaxies in the sample and the accuracy of the measured distances to the galaxies. Given the growth of photometric-only surveys, these cosmological measurements will require the use of reliable photometric redshifts and a complete understanding of their uncertainties. As a result, \pz methods will be most effective going forward if they can not only robustly provide a reliable redshift estimation but also a redshift probability density function.  

In addition, the extra information in a redshift PDF can be used to improve or enhance a particular cosmological measurement analysis. For example,~\cite{Myers2009} have shown that by using the full redshift PDF within a two-point angular quasar correlation function, as opposed to simply using a single redshift estimate, their measurement has been improved by a factor of nearly four, which is equivalent to increasing the survey volume by a similar factor. Likewise, \cite{Mandelbaum2008} discuss how the accuracy of \pz and the inclusion of the \pz PDF affect the calibration for weak lensing studies. Other recent studies~\citep[see, \eg][]{Sheth2007,van2009} have also demonstrated how a cosmological measurement can be improved by using a \pz PDF. However, given the lack of reliable \pz PDF estimation techniques, this remains an underutilized tool.

In this work, we address these issues by introducing TPZ (Trees for Photo-Z), a new, Python-based, machine learning, parallel code for estimating photometric redshift PDFs by using prediction trees and random forest techniques~\citep{Breiman1984,Breiman2001}. Our approach is an ensemble learning method that generates several classifiers and combines their results into a final output. Prediction trees partition the multi-dimensional space recursively into smaller regions, which is terminated when a leaf only contains a few elements. Within these final leaves, our algorithm can leverage a simple model for the actual prediction, by using, for example, the mean value for a regression or the mode in a voting process as used in a classification scheme.

Likewise, the basic idea of a random forest method is to use bootstrap samples from the training data to build a set of prediction trees. These trees are constructed by selecting the best split point from a random subsample of the dimensions (\eg magnitudes or colors) along which the data are subdivided. By aggregating the predictions from this forest of trees, we produce a more accurate estimate. In our implementation, we incorporate the errors on the measured attributes by perturbing the galaxy parameters by their uncertainties. We repeat this process, generating multiple individual new observations of each galaxy that are subsequently combined into a final PDF, which can be used as desired to estimate a single redshift and its associated error. In addition, our implementation of this technique naturally incorporates data with missing values and also provides extra meta information, such as an unbiased estimate of the prediction error, a measure of the relative importance of the parameters used in the \pz 
estimation as a function of redshift, an 
identification of regions where the training data provide poor predictions, and an identification of galaxies that are likely outliers.

This paper is organized as follows. In \S2 we provide a complete and detailed  description of the \pz method presented herein. \S3 introduces the different data sets we use to test the efficacy and accuracy of TPZ and its unique capabilities. In \S4 we describe the specific experiments we perform to test our \pz implementation by using these data, present an analysis of the results and discuss the capabilities of our approach. Finally in \S5, we conclude with a summary of our main results and a discussion of the TPZ algorithm.

\section{Methods}

Among the different non-linear methods that are used to compute photometric redshifts, prediction trees  are one of the simplest yet most accurate techniques. Supervised learning methods using prediction trees, either classification or regression, have been shown to be one of the most accurate algorithms for low as well high multi-dimensional data~\citep{Caruana2008}. They also are fast, can easily deal with missing data, and have similarities with other non-parametric technique. For example, prediction trees are similar to k-nearest-neighbor (kNN) algorithms in that they both group data points with similar characteristics. 

However, kNN use test data to identify similar points within the training set while keeping the parameter $k$ fixed, even though some points might have a very different number of similar neighbors. On the other hand, prediction trees have terminal leaves that bound regions of the parameter space where the predictions (\ie redshifts) and their properties (\eg  magnitudes) are similar. As both the quantity and identify of test data can vary between leaf (or terminal) nodes, prediction trees are known as \textit{adaptive} nearest-neighbor methods~\citep{Breiman1984}.

\subsection{Prediction trees\label{predtree}}
Prediction trees are built by asking a sequence of questions that recursively split the data, frequently into two branches, until a terminal leaf is created that meets a stopping criterion (\eg a minimum leaf size). The small region bounding the data in the terminal leaf node represents a specific subsample of the entire data with similar properties. Within this leaf, a model is applied that provides a fairly comprehensible prediction, especially in situations where many variables may exist that interact in a nonlinear manner as is often the case with \pz estimation. A visualization of an example tree generated by our technique is shown in Figure~\ref{fig:tree_example}. 

There are two classes of prediction trees~\citep{Breiman1984}: classification and regression, both of which are implemented in TPZ.

\begin{enumerate}[label=(\roman{*}),ref=(\roman{*})]
 \item\label{class_tree} \textit{Classification Trees (also called Decision Trees)}:
 As the name suggests, this type of prediction tree is designed to classify or predict a discrete category from the data. Each terminal leaf contains data that belongs to one or more classes. The prediction can be either a point prediction based on the mode of the classes inside that leaf or distributional by assigning probabilities for each category based on their empirically estimated relative frequencies. For example, in our \pz technique we use the magnitudes or colors of galaxies to determine the probability that a galaxy lies either inside or outside a specific redshift bin (a detailed explanation of the algorithm is presented in \S\ref{TPZal}). 
 
The tree is built by starting with a single node that encompasses the entire data, and recursively splitting the data within a node into two or more branches along the dimension that provides the most information about the desired classes. Formally this is done by choosing the attribute that maximizes the \textit{Information Gain} ($I_G$), which is defined in terms of the impurity degree index $I_d$:
 \begin{equation}\label{IG}
 I_G(T,M) = I_d(T) - \sum_{m \, \epsilon \,values(M)} \frac{|T_m|}{|T|} I_d(T_m)
 \end{equation}
where $T$ is the training data in a given node, $M$ is one of the possible dimensions (\eg magnitudes) along which the node may be split, $m$ are the possible values of a specific dimension $M$ (in the case of magnitudes $m$ might represent 2 or more magnitude bins), $|T|$ and $|T_m|$ are respectively the size of the total training data and the number of objects for a given subset $m$ within the current node, and $I_d$ is the function that represents the degree of impurity of the information. 

There are three standard methods to compute the impurity index ($I_d$). The first method is by using the \textit{information entropy}, which is defined in the expected manner (similar to Thermodynamics):
\begin{equation}\label{H}
I_d(T) \equiv H(T) = -\sum_{i=1}^{n} f_i \log_2 f_i
\end{equation}
where $i$ is the class to be predicted (\eg inside or outside a redshift bin) and the sum is over all $n$ possible classes (two in our example), and $f_i$ is the fraction of the training data belonging to class $i$. The same definition applies for a subset of the data $T_m$.

\begin{figure}
\includegraphics[width=0.44\textwidth]{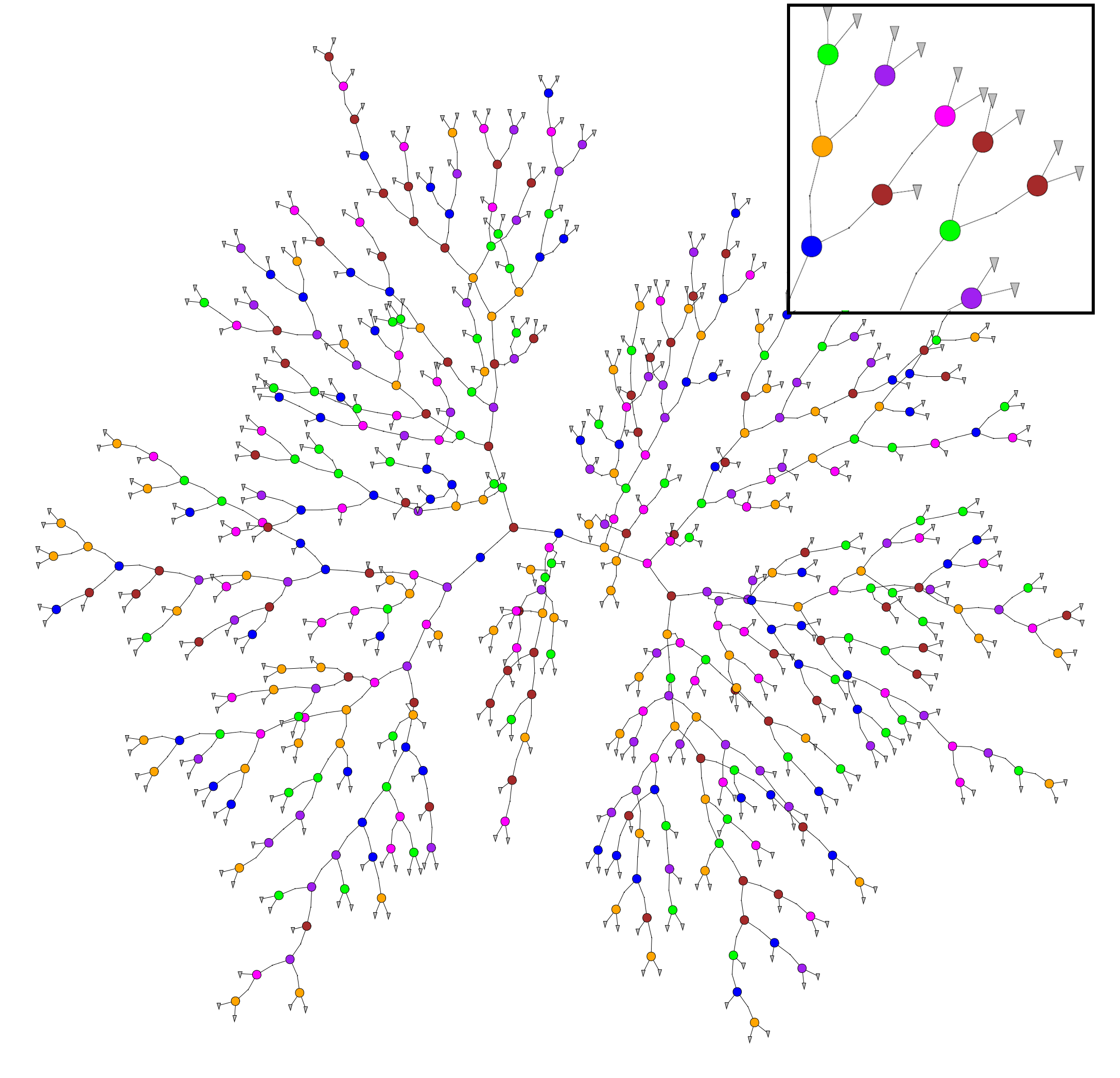}
\caption{A simplified example of a  binary prediction tree plotted radially. The initial node is close to the center of the figure. The splitting process terminates when a stopping criterion is reached. Individual colors represent the unique variable (\eg fixed aperture $g$ or $r$ or magnitude colors) used for the splitting at each node. Each leaf provides a specific prediction based on the information contained within that terminal node (gray triangles in the figure). The subpanel corresponds to zoomed in region from the tree. } 
\label{fig:tree_example}
\end{figure}

The second option, is to measure the \textit{Gini impurity} ($G$). In this case, a leaf is considered \textit{pure} if all the data contained within it have the same class. The Gini impurity can be computed inside each node:
\begin{equation}\label{Gini1}
I_d(T) \equiv G(T)=\sum_{i=1}^{n} \sum_{j\neq i} f_i f_j
\end{equation}
where $f_i$ and $f_j$ are the fractions of the training data of class $i$ or $j$. The same equation applies for a subset of $T$ along one particular dimension $M$. Since $f_i$ are the fractions for all possible classes, we have that the $\sum_i f_i =1$, and, therefore, $\sum_{j \neq i} f_j = 1-f_i$. As a result, the expression for Equation~\ref{Gini1} can be simplified to
\begin{equation}\label{Gini2}
 I_d(T) \equiv G(T) = 1 - \sum_{i=1}^n f_i^2
\end{equation}

The third method is to simply measure the impurity degree by using the \textit{classification error} ($C_E$):
\begin{equation}\label{CE}
I_d(T) \equiv C_E(T) = 1 - \max{\left\lbrace f_i \right\rbrace} 
\end{equation}
where the maximum values are taken among the fractions $f_i$ within the data $T$ that have class $i$.
During the tree construction, the data are scanned over each dimension to determine the split point that maximizes the information gain as defined by Equation~\ref{IG} and the attribute that maximizes this impurity index overall is selected. For example, Figure~\ref{fig:impurity} shows these three impurity indices, for a node with data that are only categorized into two classes, as a function of the fraction of the data having a specific class. If all of the data belong to a specific class, the impurity is zero. On the other hand, if half of the data have one class and the remaining data all belong to the other class, the impurity is at its maximum. Our implementation can calculate any of these three different impurity indices, and any one of them can be selected for the construction of the prediction trees. Alternatively, the index providing the highest information gain at a given node can be selected.

\begin{figure}
\includegraphics[width=0.44\textwidth]{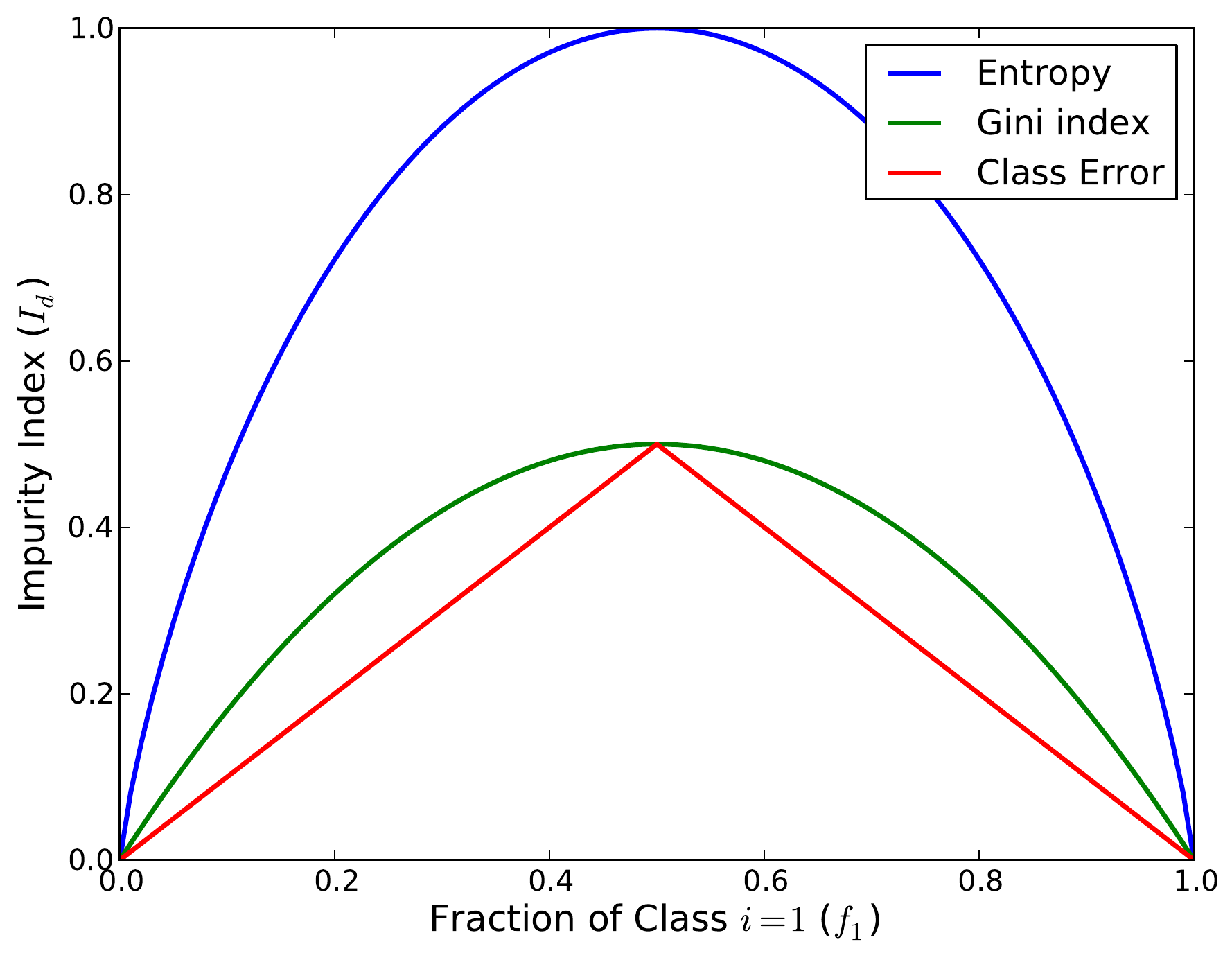}
\caption{Impurity index $I_d$ for a two-class example as a function of the probability of one of the classes $f_1$ using the information entropy(blue), Gini impurity (green) and classification error (red). In all cases, the impurity is at its maximum when the fraction of data within a node with class 1 is 0.5, and zero when all data are in the same category.} 
\label{fig:impurity}
\end{figure}

 \item\label{reg_tree} \textit{Regression Trees}
A second type of prediction tree is used when the data to be predicted is continuous; since it does not use discrete classes, we instead fit a regression model to the data inside a leaf. The construction of a regression tree follows the same structure as the classification tree, and once again a node is generally divided into two branches (\ie a binary tree). There are two primary differences, however, between regression and decision trees. First, each leaf has  training data with different redshift values; the prediction value is based on a regression model covering these points. Usually, the mean of the training redshifts is returned, so each prediction is no longer a discrete classification, but is instead an estimation of a continuous variable. Second, the procedure used to select the best dimension to split for a regression tree is based on the minimization of the sum of the squared errors, which for a node $T$ is given by
 \begin{equation}\label{S_RT1}
 S(T) = \sum_{m \, \epsilon \, values(M)} \sum_{i \, \epsilon \, m} (z_i - \hat{z}_m)^2
 \end{equation}
where $m$ are the possible values (bins) of the dimension $M$, $z_i$ are the values of the target variable on each branch/bin $m$, and $\hat{z}_m$ is the specific prediction model used. In the case of the \textit{arithmetic mean}, we have that $\hat{z}_m = \frac{1}{n_m}\sum_{i \, \epsilon \, m} z_i$, where $n_m$ are the members on branch $m$. This allows us to rewrite Equation~\ref{S_RT1} as
\begin{equation}\label{S_RT2}
 S(T) = \sum_{m \, \epsilon \, values(M)} n_m V_m
\end{equation}
where $V_m$ is the variance of the estimator $\hat{z}_m$. 

At each node in our tree, we scan all dimensions to identify the split point that minimizes $S(T)$. The splitting dimension that has the lowest value of $S$ is selected as the splitting direction, and this procedure is repeated until either some threshold in $S$ is reached or any new nodes would contain less than the predefined minimum leaf size.
\end{enumerate}

\subsection{Random forest}\label{RFs}

Random forest is an \textit{ensemble learning} algorithm that first generates many prediction trees and subsequently combines their predictions together. It is one of the most accurate empirically trained learning techniques for both low and high dimensional data~\citep{Caruana2008}. The idea is simple, given a training sample $T$ containing $N$ objects that have $M$ attributes (\eg survey magnitudes), create $N_T$ bootstrap samples of size $N$ (\ie $N$ randomly selected objects with replacement). From these samples, we create the corresponding $N_T$ prediction trees without pruning them back. 

If all the variables are examined when deciding the best point to split, the method is called \textit{bagging} \citep{Breiman1996}. An additional layer of randomness can be added to the bagging process by choosing the best split point from among a random subsample of $m_* < M$ variables at each node, where $m_*$ is kept fixed during the process. The value of $m_*$ is an adjustable parameter that is directly related to the \textit{strength} of a tree (a strong tree has a low error rate) and the \textit{correlation} between any two trees (the more correlated the trees, the higher the forest error rate). Increasing or reducing $m_*$ has the same effect on both features. Of course we want to select the optimal value of $m_*$. A good starting point is to set $m_* \simeq \sqrt{M} $, although the accuracy of the algorithm is, in the end, not very sensitive to this parameter for a large number of trees and relatively small number of dimensions. After constructing all of the prediction trees, a final and robust 
prediction is calculated by combining all $N_T$ estimates together.

\cite{Breiman2001} first introduced this algorithm and showed that this technique performs very well when compared to many other learning techniques. This technique is robust against overfitting (\ie there is no limit on the number of trees, $N_T$, in the forest), it runs efficiently on large data sets, it can generate an internal unbiased estimate of the error, and it can provide extra information about the relative importance of the input variables and the internal structure of the training data.

\subsubsection{Ancillary information}\label{AI}

Given a training set $T$, this extra, ancillary information can be calculated prior to the computation of the \pz PDFs. As a result, we can use this {\em a priori} information to explore the efficacy of different parameter combinations while also obtaining an estimate of the bias and variance of the \pz prediction. This is done by using \textit{out-of-bag} (OOB) samples, which consist of a random sample of data that are left out of each tree. In the process of growing a forest, $N_T$ trees are created using bootstrap samples of size $N$. In each of these samples, about one-third of the data are not used when constructing a tree, and are instead used as a test sample for the recently built tree. The test results created by using this OOB data are combined together to obtain estimators of the error, which, when built using a sufficiently large number of trees in the forest, has been shown to be unbiased and as accurate as using a validation set of the same size as the training set~\citep{Breiman1996}. This 
removes, 
therefore, the need 
for a separate validation sample that can introduce a bias into the final result. This method also has the advantage of using the full spectroscopic data to compute PDFs.

The OOB data can also be used to estimate the relative importance of each attribute or dimension to the \pz calculation. This provides an elegant method to identify and remove attributes that do not contribute significantly, thereby reducing the noise and  dimensions of the problem. This also has the benefits of increasing the performance of the implementation, improving our understanding of the complexity in the interaction between different attributes, and improving the identification of new training data from, for example, follow-up observations. This relative importance is estimated for each attribute by first quantifying any variations in the prediction error when the OOB data are permuted only along the specific attribute, leaving the others unchanged. This process is repeated across all trees, and the end result is the average in the error increment when compared to the unperturbed variables for all the tress over the entire forest.

Another item we can construct is the proximity matrix, $Prox(i,j)$, which is a symmetric, positive definite matrix that gives the fraction of trees in the forest in which element $i$ and $j$ fall in the same terminal leaf. This matrix is constructed tree-by-tree by running all the data, both the OOB and the data used for growing, down each tree. When galaxy $i$ and $j$ are in the same leaf, their proximity is increased by one. At the end, all the proximities are normalized by the total number of trees; therefore, similar galaxies will tend to have higher proximities than dissimilar ones. This matrix can be computed for the training set, the test set, or both together. Since this matrix quantifies the relative similarity between galaxies, it can be used to identify outliers within a data set. For example, by computing the squared sum of all proximities for each galaxy, we can algorithmically identify galaxies with few neighbors by selecting sources with the lowest value, which can be flagged for further 
inspection. 

To build or apply prediction trees, the data cannot have missing values for any of the attributes used to construct the trees (\eg the most important survey magnitudes). To include more data into the classification process, we can use the proximity matrix to estimate any missing values or to replace highly uncertain values. We do this in an iterative process, by performing the forest growing step of the algorithm and replacing the missing attribute at each pass. We select the replacement value by computing the average parameter value from the $k$ nearest galaxies; we can also inversely weight these galaxies by their respective distance. This process continues until we have obtained convergence or until a fixed number of iterations have been performed.

By using the proximity matrix, OOB error estimates, and the relevant importance of different attributes, we can also identify zones where the \pz prediction is either poor or is loosely constrained by the training data. In either case, this knowledge is of vital importance when deciding what galaxies to target spectroscopically in order to optimally improve a training sample. One way this feature is implemented is by using the two most important attributes to map the areas of parameter space by their prediction error. This map can guide the identification of new data that increases the efficacy of the training sample by targeting those galaxies that minimize the prediction error in under sampled areas, thereby more effectively utilizing limited spectroscopic follow-up observations.

\begin{figure*}
\includegraphics[width=0.88\textwidth,height=0.55\textwidth]{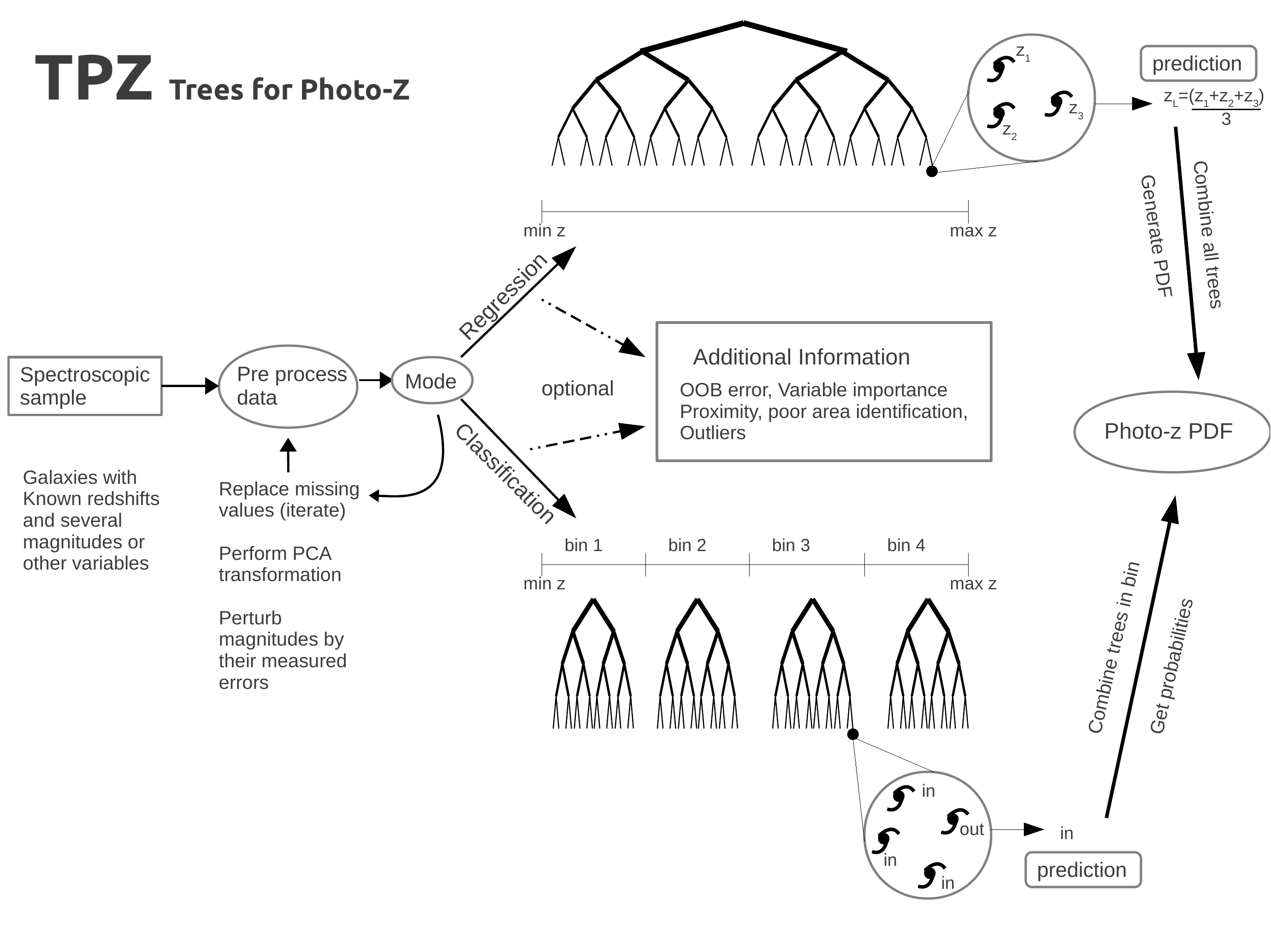}
\caption{A simplified representation of TPZ, the details for each subprocess are described more fully within the text. Note that each tree drawn in the figure represents a full random forest with $N_T$ bootstrap samples for every one of the $N_R$ random perturbed samples. The big circles containing galaxies represent a terminal leaf, which are directly used to make a prediction for each new galaxy.} 
\label{fig:TPZ}
\end{figure*}

\subsection{Previous work}
Two previous works have utilized prediction trees for \pz calculations.~\cite{Carliles2008,Carliles2010} predicted photometric redshifts and their errors by using a random forest built with the regression tree package for $R$ by using the mean value as their leaf model. They used a subset of the main galaxy sample of the SDSS Data Release 6~\citep{Adelman-McCarthy2008} catalog with colors as their attributes. They demonstrated that random forest methods are well suited to the \pz estimation problem as they obtained comparable results to other machine learning methods, and they publicly released their $R$ scripts. They did not, however, take full advantage of the ancillary information provided by the random forest technique, nor did they produce probability density functions. 

\cite{Gerdes2010} have developed a new technique, called ArborZ, to compute photometric redshifts using boosted decision trees (BDT). These classification trees are constructed in a similar manner to our classification trees, as discussed in \S\ref{predtree}. In their approach, all data points start with equal weights, but after each tree is built, higher weights are assigned to points that were previously misclassified. This process iteratively combines weak classifiers into a single stronger one~\citep{Schapire1998}; and, in the end, a weighted vote across the classifiers produces the final prediction. In their approach, they divide the redshift range into small bins and use an ensemble of BDTs to generate a probability distribution. A photometric redshift is estimated by determining the mean value of this distribution. They tested this algorithm on SDSS DR6 data as well as DES simulated data, finding similar performance to other empirical training methods, such as the \pz estimates provided by~\cite{
Oyaizu2008a} in the 
case of the SDSS data, and by ANNz~\citep{Collister2004} for the DES. 

Our approach, detailed below, extends these previous results to create a new publicly available method that uses random forests to compute PDFs by using classification and/or regression trees. Our approach also uses extra information encoded within the measurement errors, generates extra, ancillary information describing the spectroscopic training sample, and provides a better control of the uncertainties. We also, therefore, are able to examine the importance of the attributes used to grow the trees, and identify areas in the attribute space where the training data are dominated by shot noise statistics.

\subsection{TPZ Algorithm}\label{TPZal}

Our implementation of prediction trees with random forest for photometric redshift PDF prediction, TPZ, is written in the Python\footnote{http://www.python.org/} programming language and uses MPI for parallel communication to run efficiently on distributed memory systems. As shown in Figure~\ref{fig:TPZ}, our implementation is divided into three steps:

\begin{description}
\item[Data Pre-processing]  The \textit{first step} prepares the data for the construction of the prediction trees. First, we optionally perform a principal component analysis (PCA) of the data in order to reduce strong correlations between attributes. This PCA transformation can reduce the dimensionality of the input data prior to the training, which can be important for large data sets with many attributes. This step also includes the replacement of missing values, which we do iteratively, finding that between $5$--$10$ iterations leads to a convergence on the missing values. We next generate $N_R$ training samples by perturbing the measured values according to the error on each variable, which we assume to be normally distributed. In this manner, we can incorporate the measurement error in the prediction tree construction, we reduce the bias on proximity 
matrices, and we introduce randomness into the construction of the trees in a controlled manner.

\item[Random Forest Construction]  The \textit{second step} is the actual construction of the random forest, where we generate fully grown prediction trees. We construct $N_T$ trees by using bootstrapping for each perturbed sample in the set of $N_R$ training samples we created in the first step. This step can be done several times with a smaller number of trees to both explore the parameter space and gain insight into the internal structure of the data prior to building the final prediction trees. Finally, this step can also produce the ancillary information that can characterize the performance of our code prior to estimating the final \pz values.

\item[\PZ PDF Construction] The \textit{final step} uses the newly generated prediction trees to create individual \pz PDFs for each source in the application data set. This process involves running each source down each tree, testing the source at each node until we arrive at a terminal leaf where we make a prediction. At the end, we combine all of the forest predictions into a probability density function.
\end{description}

\subsubsection{Implementation modes}\label{Implement}
TPZ can use either type of prediction tree that uses random forests: classification or regression; the actual implementation details only differ after the first step.
\begin{description}
 \item[Classification Mode:] In this mode, the spectroscopic sample is divided into several redshift bins that either have a fixed width (or, alternatively, resolution), which allows a variable number of galaxies within each redshift bin, or have a fixed number of galaxies per redshift bin, which means our redshift bins are of variable width. Within each bin, we create a forest of classification trees, as described above, using the perturbed samples as well as the bootstrap samples.
 These trees classify an object as either lying \textit{inside} or \textit{outside} a bin. By using all of the training data within each bin, we  both decrease the overall performance of our implementation due to the larger data volume and also increase the chance of catastrophic errors since most data will lie outside the bin of interest.

We address these issues by following a similar approach to that used by \cite{Gerdes2010}. For each bin, we identify all  sources that lie inside the bin. This number of galaxies with class \textit{inside} is $n_{in}$. We next select a factor $fn_{out}$ of the $n_{in}$ galaxies that have spectroscopic redshifts that lie outside the bin by a factor of $z_{out}$ times the width $\delta z$ of the bin. This means that galaxies with class \textit{outside} fall $z_{out} \times \delta z$ from the boundaries of the bin. This allows a better distinction between the class \textit{inside} and the class \textit{outside} as it would have if we include objects located very near to these boundaries. In the end, each bin will have $(1+fn_{out})n_{in}$ galaxies available for training the forest. 

If the training set is limited, wider bins can be used in order to have a sufficient number of training galaxies per bin. Furthermore, these bins can even be allowed to overlap by some value; this overlap can be taken into account when building the \pz PDFs by normalizing by the fraction of wider bins that overlap with each other.  After all of the forests are created for all of the bins, the test data are run down each tree in each forest, which assigns either the class \textit{inside} or \textit{outside} to the test source. After combining all of the assigned classes from the forest, we assign a probability for the source to belong to that redshift bin, which is simply the number of times the source was assigned the \textit{inside} class divided by the total number of trees. By repeating this process for each bin and renormalizing the subsequent result, we generate a \pz 
PDF for the source.

 \item[Regression Mode:] In this mode, we use all available training data to fully grow each tree. For each perturbed sample, $N_T$ trees are created using the methodology explained in \S\ref{predtree}\ref{reg_tree}. At the end, there is one large random forest covering the entire spectroscopic range. The difference with the classification mode is that, after the tree has been constructed by splitting the nodes according to Equation~\ref{S_RT2}, each terminal leaf only ends up with a few sources to make the prediction. In the simple case of obtaining a single estimate, this leaf can be replaced by the mean or the median of the values inside it; more generally, these values are kept for computing the PDF. To compute a \pzns, the test data are run down each tree in the forest. Each tree returns the set of spectroscopic redshift measurements that, after conversion to a given resolution, are converted into a PDF by normalizing to the total number of objects returned. All  trees 
have the same weight when constructing the PDF, as well as the values of the terminal leaves identified in each tree. If a single value is desired, a mean value and its error can be returned via the standard methods by aggregating all of the relevant values as returned by the different trees.
\end{description}

The choice of either of these modes will depend on the characteristics of the data being analyzed. On average, the regression mode runs faster than the classification mode for a specific accuracy, and is also better suited for data that are not uniformly distributed. The classification mode, on the other hand, provides a better characterization of the data as a function of redshift, since it creates its own random forest on each bin unlike the regression mode where a forest is created using the full range in redshift. The classification mode is also better suited for uniformly distributed data and can provide a reliable and robust prior probabilities in a Bayesian framework when using wider redshift bins. When faced with a high quality and rich training set, both modes will provide similar accuracies and error rates, but the regression mode, being faster, would generally be preferred.

Figure~\ref{fig:TPZ} shows a simplified workflow of our TPZ implementation. 
Each tree in this figure represents an entire forest, where the single tree results are averaged to get a final prediction. The classification mode predicts a probability that a source lies within each bin, thereby building up a \pz PDF, while the regression mode keeps all sources found on a terminal leaf and combines their values to construct a \pz PDF at the desired resolution. For both modes, ancillary information can be provided, and both modes share the same data pre-processing steps.

\begin{figure*}
\includegraphics[width=0.33\textwidth,height=0.3\textwidth]{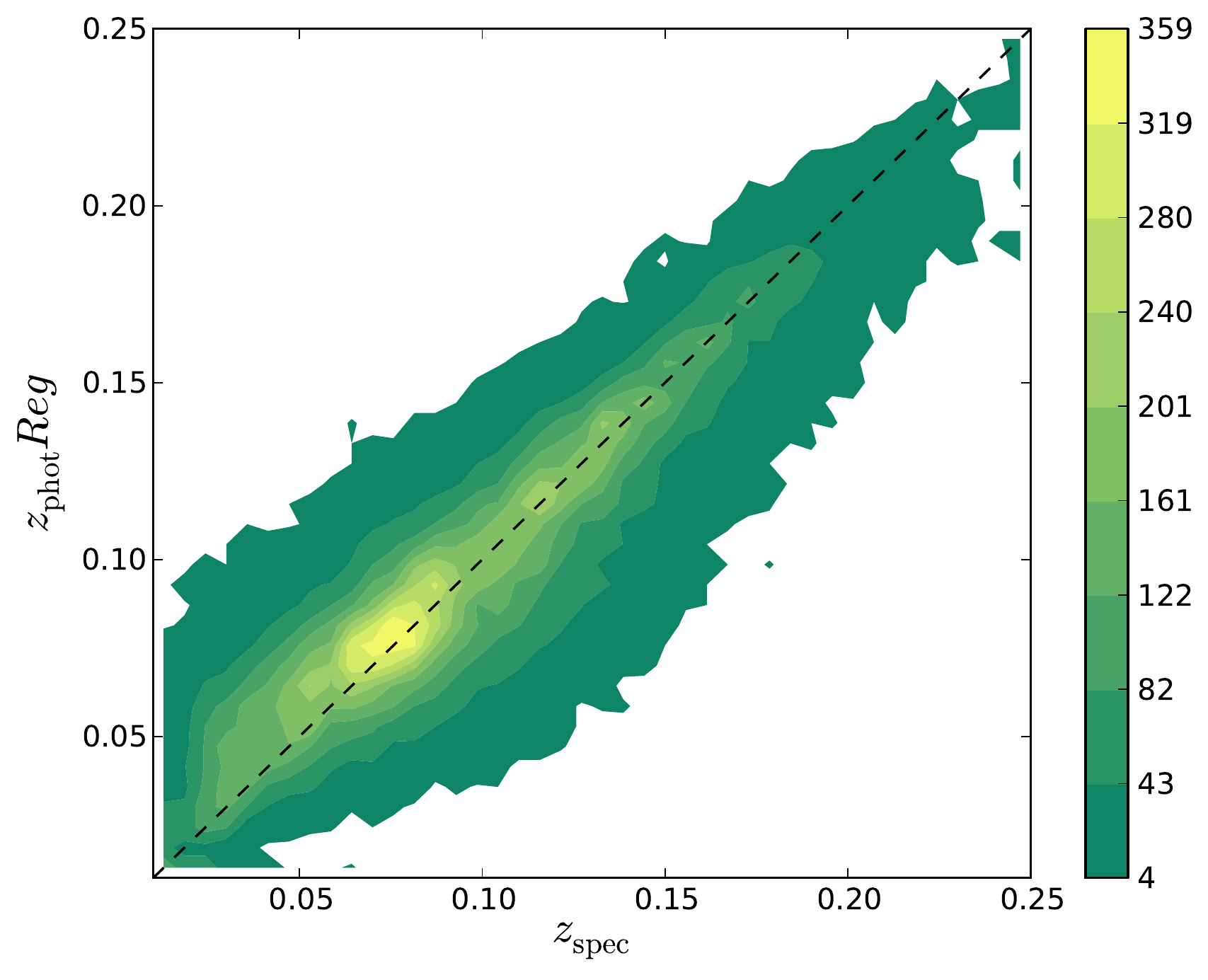}
\includegraphics[width=0.33\textwidth,height=0.3\textwidth]{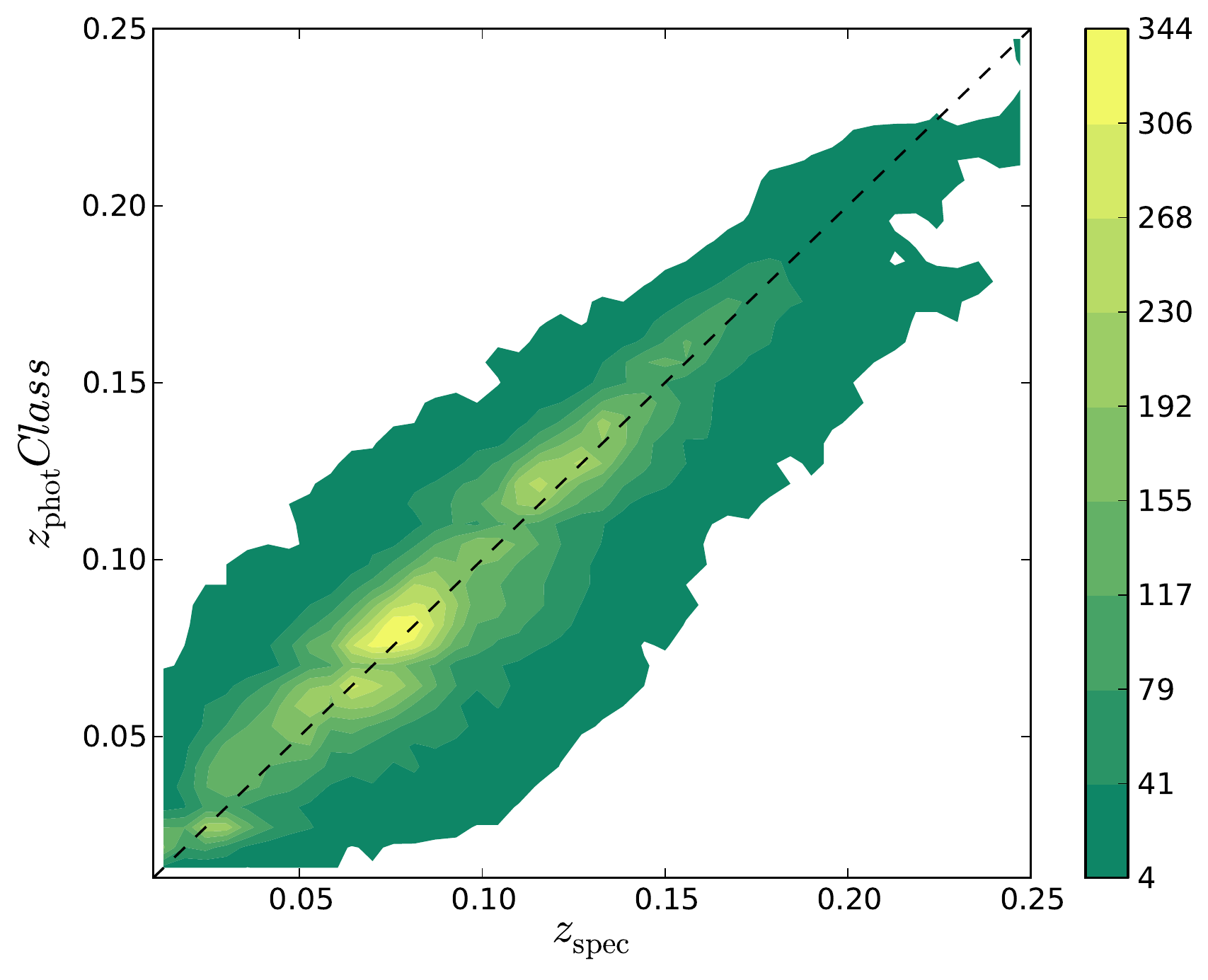}
\includegraphics[width=0.33\textwidth,height=0.3\textwidth]{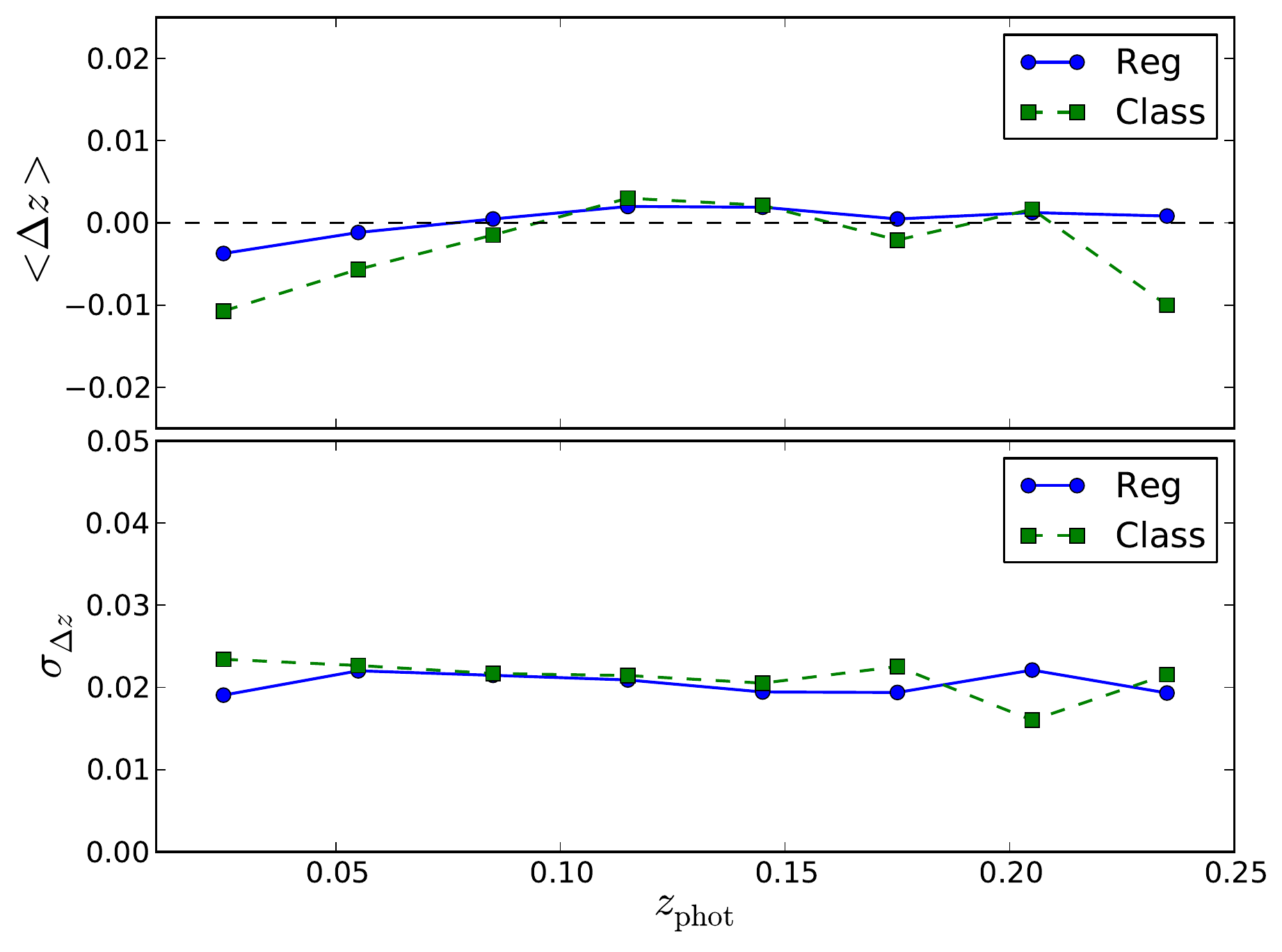}
\caption{Photometric vs. spectroscopic redshift for all test SDSS MGS test galaxies using  regression mode (\textit{Left}) and classification mode (\textit{Center}). (\textit{Right}) A comparison of the bias (upper panel) and the scatter (lower panel) as a function of redshift for the SDSS MGS data by using the regression mode (blue dots) and the classification mode (green squares). }
\label{fig:true_plot}
\end{figure*}

\section{DATA SELECTION}\label{pzdata}

To demonstrate the capabilities and the efficacy of the different parameter configurations of the TPZ code, we have used several different photometric and spectroscopic data sets that vary both in quantity and quality. In this section we briefly discuss these  data sets and the specific data samples from each that we used in the testing process, which is further described in \S\ref{App}.

\subsection{Sloan Digital Sky Survey}
The Sloan Digital Sky Survey~\citep[SDSS;][]{York2000} phase I and phase II conducted a photometric survey in the optical bands $u$, $g$, $r$, $i$, $z$ that covered almost 10,000 square degrees, or approximately one-fourth of the entire sky. The resultant photometric catalog contains photometry of over $10^8$ galaxies, making the SDSS one of the largest surveys ever performed. The SDSS also conducted a spectroscopic survey of targets selected from the SDSS photometric catalog, obtaining spectra of about $10^6$ low redshift galaxies. 

In this paper, we use a subset of the Main Galaxy Sample~\citep[MGS;][]{Strauss2002} from the Data Release 7 catalog~\citep{Abazajian2009}. Specifically, we selected 55,000 galaxies by using the online CasJobs website\footnote{http://casjobs.sdss.org/CasJobs/}. This spectroscopic data ranges from $z \approx 0.02$ up to $z \approx 0.3$ with a mean redshift of 0.1. From this sample, we randomly selected 15,000 galaxies to train the TPZ implementation, while holding the remaining 40,000 for testing. We note that this is a blind test, as the testing data are not used in any way to train or calibrate the TPZ algorithm. Of all the measured attributes in the SDSS photometric catalog, we have used only the four dimensions corresponding to the galaxy colors as derived by the extinction corrected model magnitudes : $u - g$, $g - r$, $r - i$, and $i - z$. We use the SDSS colors as opposed to the more commonly used magnitudes for this particular test to both demonstrate the flexibility of TPZ and to generate 
scientifically more interesting ancillary information.

\subsection{PHoto-z Accuracy Testing Project}\label{phat1data}
The PHoto-z Accuracy Testing~\citep[PHAT;][]{Hildebrandt2010} project first compared the performance and systematics of different \pz codes on synthetic data (PHAT0) that was specifically created for a contest, and also more recently used real data (PHAT1) in a similar manner; thereby providing a more realistic comparison by using real measurements. The PHAT project\footnote{www.astro.caltech.edu/twiki\_phat/bin/view/Main/WebHome} provides filter responses for \pz estimation by SED-fitting methods and a training data set for \pz estimation by empirical methods. The true redshifts of the test data are not public, which provides a more reliable, blind comparison between different approaches~\citep[see][for more details about the contest]{Hildebrandt2010}. In this paper, we use the PHAT1 data, which consists of real observations selected from the Great Observatories Origins Deep Survey Northern field~\citep[GOODS-N;][]{Giavalisco2004}.

These data include photometry from the original ACS four-band data: F435W(B), F606W(V+R), F775W(i') and F850LP(z') that have been cross-matched with photometry from~\cite{Capak2004}, including $U$ (from KPNO), $B_J$, $V_J$, $R_C$, $I_C$, $z^{'}$ (from SUBARU), and $HK^{'}$ (from QUIRC). In addition, the photometry of PHAT1 also includes Deep $J$ and $H$ bands~\citep[from ULBCAM;][]{Wang2006}, $K_S$~\citep[from WIRC;][]{Bundy2005}, and four Spitzer IRAC bands: $3.6$, $4.5$, $5.8$, and $8.0 \micron$. This photometric catalog was cross-matched with all available spectroscopic GOODS-N data~\citep{Cowie2004,Wirth2004,Treu2005,Reddy2006}, producing a final data set of eighteen band photometry and spectroscopy for 1,984 galaxies. 

For the contest, only 515 galaxy redshifts were published for use as training data; the remaining redshifts were unpublished and used internally by the PHAT project to conduct a blind comparison test. Despite the limited training data, multiple authors submitted the \pz predictions and the results were published in \cite{Hildebrandt2010}. As the contest had already been completed when we started this work, we were unable to participate. However, as discussed in \S\ref{App} we have tested TPZ on the PHAT1 training data in an analogous manner as the contest and have submitted our results to the official PHAT wiki.

\subsection{Deep Extragalactic Evolutionary Probe }\label{deep2data}
The Deep Extragalactic Evolutionary Probe (DEEP) is a multi-phase, deep spectroscopic survey performed with the Keck telescope. Phase I used the LIRS instrument~\citep{Oke1995}, while phase II used the DEIMOS spectrograph~\citep{Faber2003}. The
DEEP2 Galaxy Redshift Survey is a magnitude limited spectroscopic survey of objects with $R_{AB} < 24.1$~\citep{Davis2003,Newman2012}. The survey includes photometry in three bands from the CFHT 12K: $B$, $R$, and $I$ and it has been recently extended by cross-matching the data to other photometry databases. In this work, we use the Data Release 4~\citep{Matthews2013}, the latest DEEP2 release that includes secure and accurate spectroscopy for over 38,000 sources. The photometry for the sources in this catalog was expanded by using two $u$, $g$, $r$, $i$, and $z$ surveys: the Canada-France-Hawaii Legacy Survey~\citep[CFHTLS][]{Gwyn2012}, and the SDSS. For additional details about the photometric extension of the DEEP2 catalog see~\cite{Matthews2013}. 

To use the DEEP2 data with TPZ, we selected sources with secure redshifts (ZQUALITY$\geq 3$) that were securely classified as galaxies, have no bad flags, and have full photometry. Even though the filter responses are similar, the $u$, $g$, $r$, $i$, and $z$ photometry come from two different surveys and are thus not identical. We therefore treat those galaxies with SDSS photometry for fields 2,3, and 4 of the DEEP2 target areas independently from those for field 1 with CFHTLS photometry. In the end, this leaves us with a total of 19,699 galaxies with eight band photometry and redshifts, of which we use 10,000 for training and hold the rest for testing.

\section{Application/Discussion}\label{App}

In this section, we apply the \pz estimation technique presented in \S\ref{TPZal} to the SDSS main galaxy sample (MGS), the PHAT1 blind test sample, and the DEEP2 sample, which were all introduced in \S\ref{pzdata}. Since the point of this paper is to introduce the TPZ algorithm and our associated implementation, we use these three different data sets to highlight different features of the code. Thus we do not apply TPZ uniformly to each data set, and the three subsections herein are necessarily different.


\subsection{SDSS Main Galaxy Sample}\label{SDSS-MGS}

\begin{table}
\begin{minipage}[]{\columnwidth}
\caption{A comparison between the Regression Mode and the Classification mode for the SDSS MGS galaxies with different confidence level restrictions.}
\label{tab:Reg_Class}
\centering
\renewcommand{\footnoterule}{}
\begin{tabular}{|l|rrrr|}
\hline
\hline
 Implementation &  $< \Delta z > [10^{-3}]$ & $\sigma_{\Delta z} [10^{-2}]$ & Fraction\footnote{Fraction of galaxies remaining after a cut on $zConf$.}\\
\hline
Reg All          & $-0.08$ & $2.25$ & 100\%  \\
Class All          & $2.18$ & $2.46$ & 100\%  \\
\hline
Reg  $zConf > 0.6$ & $-0.20$ & $2.18$ & 98.2\%  \\
Class $zConf > 0.6$ & $-2.15$ & $2.34$ & 94.2\%  \\
\hline
Reg  $zConf > 0.75$ & $-0.33$ & $1.97$ & 91.0\%  \\
Class $zConf > 0.75$ & $-1.80$ & $2.20$ & 73.5\%  \\
\hline
Reg  $zConf > 0.9$ & $-0.23$ & $1.76$ & 67.3\%  \\
Class $zConf > 0.9$ & $-0.92$ & $1.82$ & 34.7\%  \\
\hline
\end{tabular}
\end{minipage}
\end{table}

We first apply TPZ to the SDSS main galaxy sample, using both the regression and the classification methods as explained in \S\ref{Implement}, and we present the results in Figure~\ref{fig:true_plot}. The left and center panels compare the estimated photometric redshifts to the spectroscopic redshifts for all 40,000 galaxies held out for testing from the SDSS MGS, for regression and classification modes, respectively. Both implementations show similar performance in the central part of the redshift distribution; however, there are differences at both the low and high redshift regions of this sample. The right panel shows both the mean of the bias, defined as $\Delta z = z_{\rm spec}-z_{\rm zphot}$, and its scatter for eight redshift bins. The regression mode performs slightly better at all redshift bins, but especially on the first and last bin, where the classification mode shows systematic errors in classification. 

This error arises due to the lack of training data at those redshifts for the classification mode, where, though we allow some overlap between bins, we keep the bin size constant, which can result in large differences in the number of training objects per bin. This reduction is most pronounced in the lowest and highest redshift bins, which results in a lower accuracy and a higher scatter. We also are affected at the low redshift regime by the fact that a predicted redshift can not be negative, those introducing a positive skew to the predicted redshift values for very low redshifts.

\begin{figure}
\includegraphics[width=0.46\textwidth,height=0.41\textwidth]{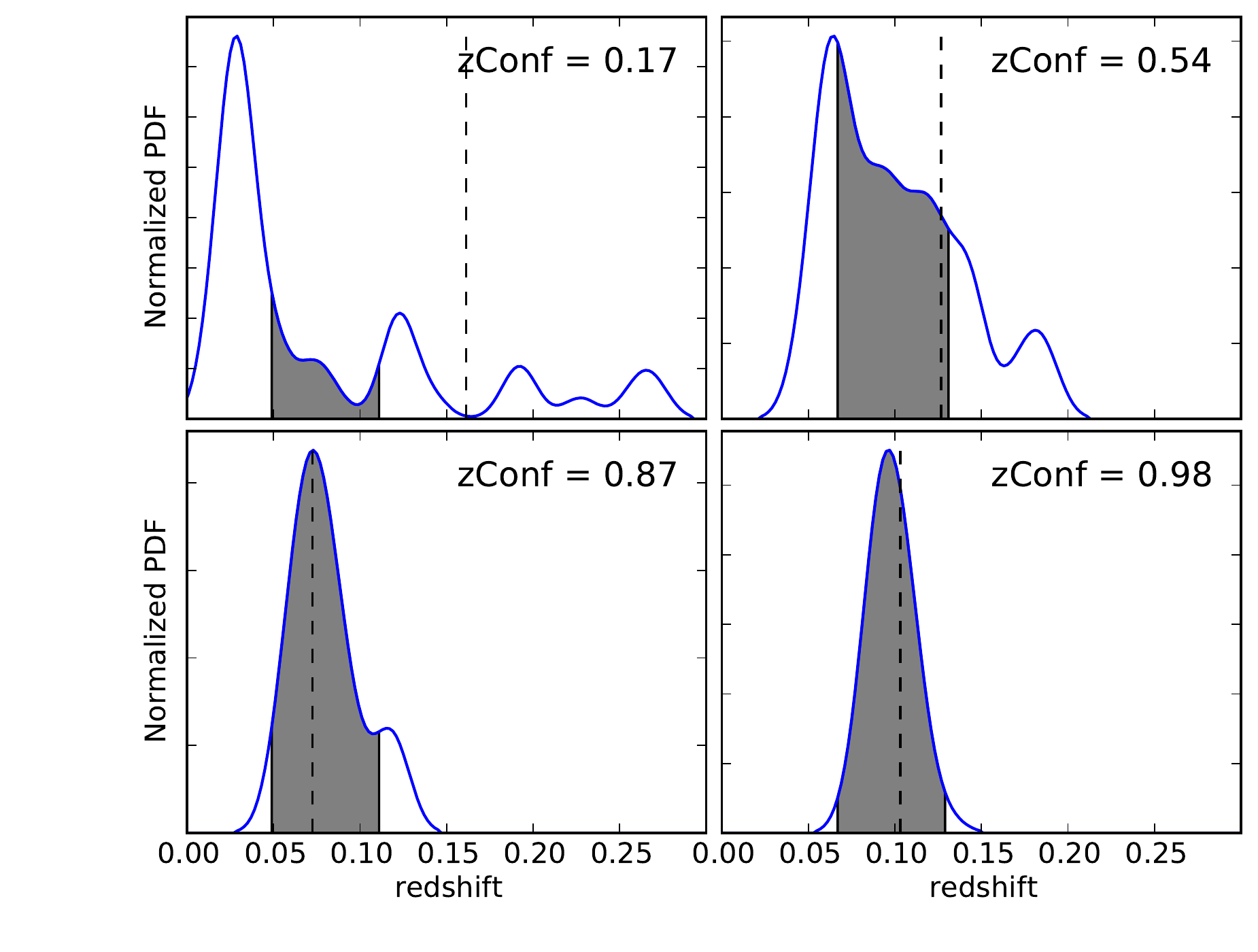}
\caption{Four example PDFs produced by TPZ for the SDSS MGS selected with different values of $zConf$. The higher the value of $zConf$, the more narrowly concentrated the PDF is about the mean.  The vertical dashed line corresponds to the spectroscopic value for the test galaxy and the gray area encloses the confidence level.} 
\label{fig:pdf_ex}
\end{figure}

Since both implementation modes produce \pz PDFs, we can compute confidence levels, $zConf$, around the mean (or mode) for each individual PDF. To simplify comparisons with past results, we define $zConf$ as the integrated probability between $z_{\rm phot} \pm \sigma_{TPZ}(1+z_{\rm phot})$. We select $\sigma_{TPZ}=0.03$ as an approximation to the intrinsic scatter of the algorithm when applied to the data, which can be computed by using the OOB data. Of course we could define $zConf$ in some other manner, but the results would be relatively unaffected.  Figure~\ref{fig:pdf_ex} presents four different PDFs taken from the SDSS MGS, each with different confidence levels that are shown as a bounded gray area under each PDF curve. 

In this example, we measured $zConf$ around the mean of each PDF and the actual spectroscopic redshifts are shown as vertical dashed lines for reference. From this figure, we see that $zConf$ provides a reasonable summary of the concentration of the PDF, and can, therefore, be used to further restrict a \pz sample by selecting only those PDFs with a $zConf$ value above some threshold. In general, as shown in this Figure, we see that lower confidence values are strongly correlated with less accurate predictions. Nevertheless, it is still possible to have a small fraction of galaxies with high $zConf$ PDFs that are estimated at the wrong redshift. We discuss the $zConf$ parameter and its use in identifying a clean galaxy sample in further detail in \S\ref{DEEP2pz}.

In Table~\ref{tab:Reg_Class}, we present the mean value of the different performance metrics described in the previous paragraphs, as applied to the SDSS MGS, as well as the fraction of remain galaxies that remain in the sample after a cut on $zConf$. As before, we see that, on average, the regression mode outperforms the classification mode on this data set, although the difference is reduced when we apply a cut on the confidence level. Interestingly, at more restrictive $zConf$ cuts, the performance of both modes is similar; however, the number of galaxies remaining in the regression mode sample is higher. Note that since these are averaged values over the sample, any minor change implies a significant change on individual calculations.

As a result, we believe that making a cut on $zConf$ results in a cleaner sample, as shown by the improved performance metrics for either implementation mode. The difference in the fraction of galaxies that remain in each sample indicates that, on average, PDFs generated by the classification implementation are broader than PDFs  generated by the regression implementation. This result is reasonable, as the classification mode bins the redshift space and provides probabilities for all bins which can produce a more sparse distribution. In the classification mode the probabilities are computed individually for each redshift bin, which could be important and easily extended to build a prior distribution that can be used in a Bayesian method. Since the regression mode was shown to be more accurate for the SDSS (see, e.g., Figure~\ref{fig:true_plot} and Table~\ref{tab:Reg_Class}), we use the mean of the PDF as calculated by the regression mode on the SDSS MGS data in the rest of this section, unless otherwise 
indicated.

We can broadly compare our use of $zConf$ to define clean galaxy samples to other published results; we note that a direct, one-to-one, comparison is problematic due to the different training sets and attributes used in computing photometric redshifts for the SDSS main galaxy sample. If we take a $zConf >= 0.75$, we keep 91\% of the data and compute the fraction of galaxies with $|\Delta z| < z_i$, where $z_i = $ 0.001, 0.002 and 0.003 as 45.2\%, 73.0\% and 89.8\%, respectively. These valued compare favorably  to those from ~\cite{Laurino2011} who, even though they used an extended catalog, compute these same values to be 43.4 \%, 72.4\% and 86.9\%, with a mean bias of $<\Delta z> = 15 \times 10^{-3}$ and $\sigma_{\Delta z} = 1.52 \times 10^{-2}$ (these latter values can be compared with our results shown in Table~\ref{tab:Reg_Class}). Finally, we note that making a strict cut of  $\Delta z > 0.006$ identifies an outlier fraction of 1.54\%, while other groups, using extended catalogs as well, have reported 
values of 1.
9\%~\citep{Gerdes2010} and 2.6\%~\citep{Oyaizu2008a}.

\subsubsection{Ancillary information}

\begin{figure}
\includegraphics[width=0.45\textwidth,height=0.41\textwidth]{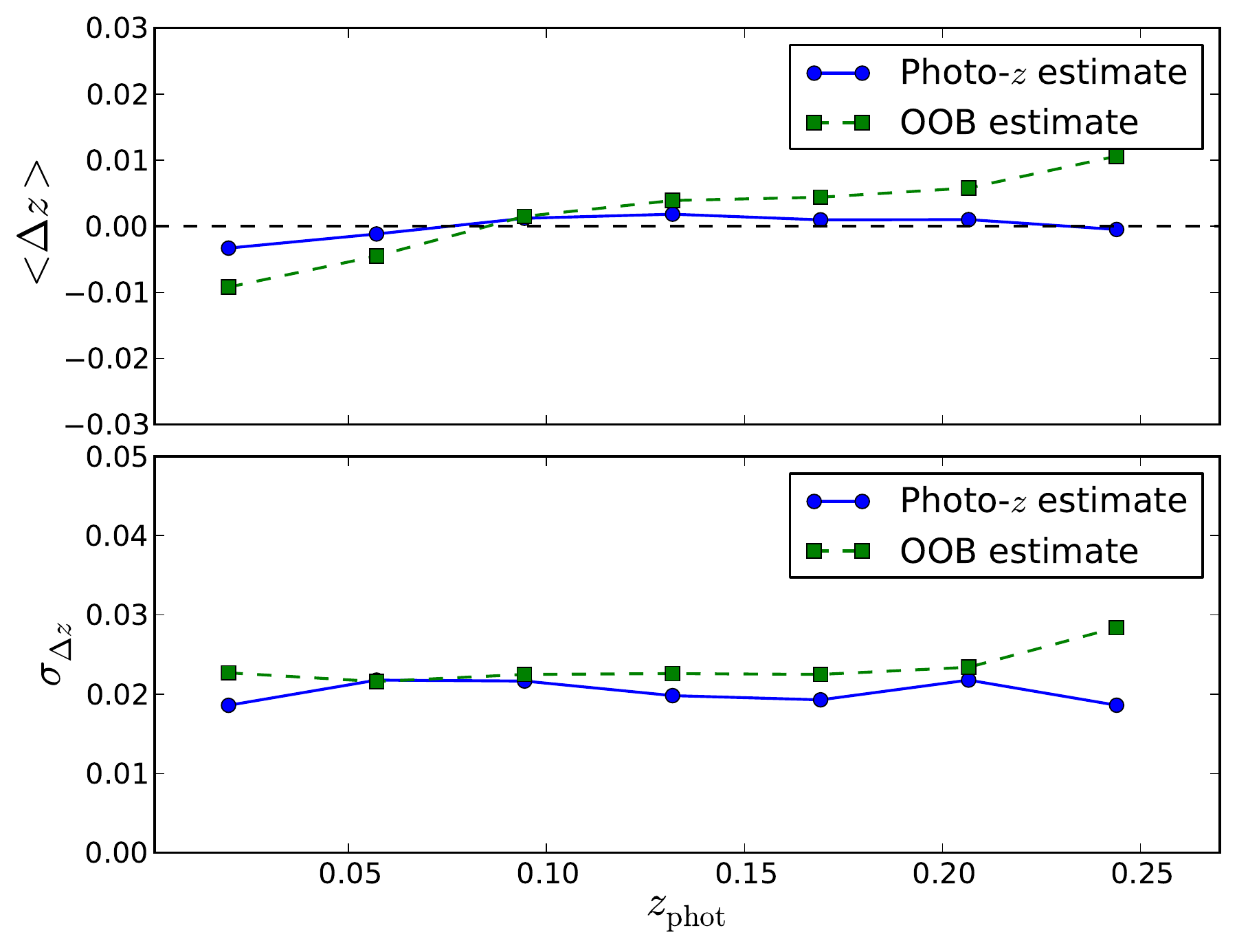}
\caption{(\textit{Top}) The averaged $\Delta z$ as a function of redshift for all test galaxies from the SDSS MGS (blue circles) and from the OOB (Out-Of-Bag) data computed individually for each tree and subsequently averaged over the forest (green squares). (\textit{Bottom}) The standard deviation of $\Delta z$ as a function of redshift for the test set (blue circles) and the OOB data (green squares). In this case the OOB data provide a unbiased, upper-limit for these metrics.} 
\label{fig:oob}
\end{figure}

As detailed in \S\ref{AI}, we can use the out-of-bag data to compute extra, ancillary information about the SDSS MGS dataset. For this purpose, we first select approximately one-third of the objects from each bootstrap sample. Using these data, we compute an unbiased indicator of the bias (\ie $\Delta z$) and its standard deviation (\ie $\sigma_{\Delta z}$) for each tree. Finally, we average these metrics over all trees. In Figure~\ref{fig:oob}, we present in the top panel the mean bias as a function of redshift taken both from the test data (blue line) and from the OOB data used during the training process (green line). The bottom panel in this figure presents the standard deviation for each redshift bin. 
The RMS of these values provides an approximation to the intrinsic error and scatter of the TPZ code, which can be used to compute  confidence levels. From the OOB data, we compute the RMS of the bias to be $0.0064$, which can be compared to the value of $0.0017$ obtained directly from TPZ for the SDSS MGS test sample. Likewise, we can approximate the scatter; for the OOB data we have $0.0235$, while for the SDSS MGS test sample we have $0.0203$. Thus, the OOB data provide upper limits for these metrics calculated by using only the training sample.

This OOB technique is unique due to the fact that the OOB data were not used to train a particular tree, yet the full data are used when building the forest by using the bootstrap samples.  If we would have run all of the training data \textit{after} the forest was constructed without using the OOB approach, we would have obtained biased (although lower values) for these metrics. This approach would thus not provide a prior estimation of the accuracy of TPZ. With the OOB data, we compute \textit{a priori} these unbiased estimates exclusively from the training set, without the need for a validation set, allowing us to take full advantage of all available spectroscopic data.

\begin{figure*}
\includegraphics[width=0.325\textwidth,height=0.3\textwidth]{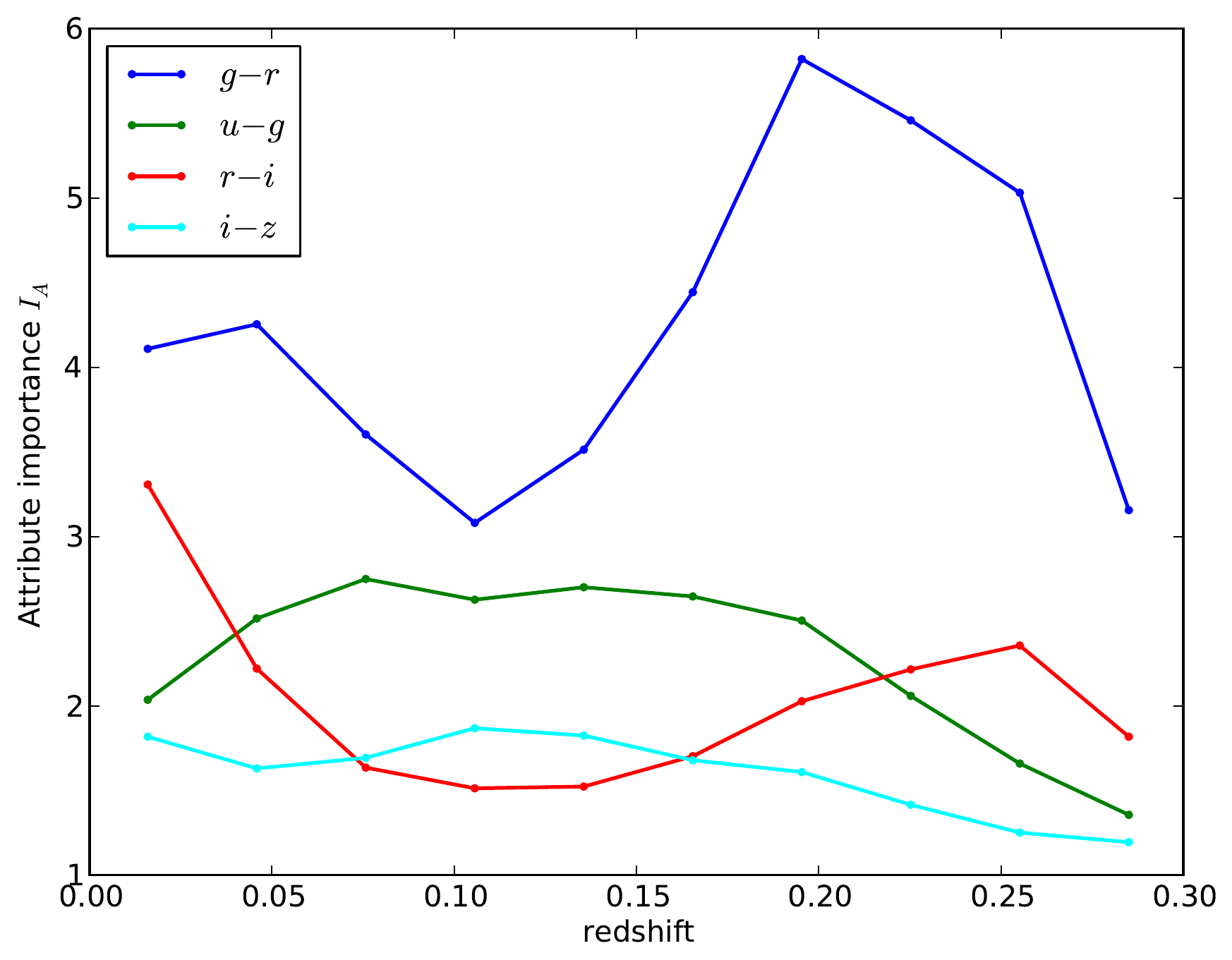}
\includegraphics[width=0.325\textwidth,height=0.3\textwidth]{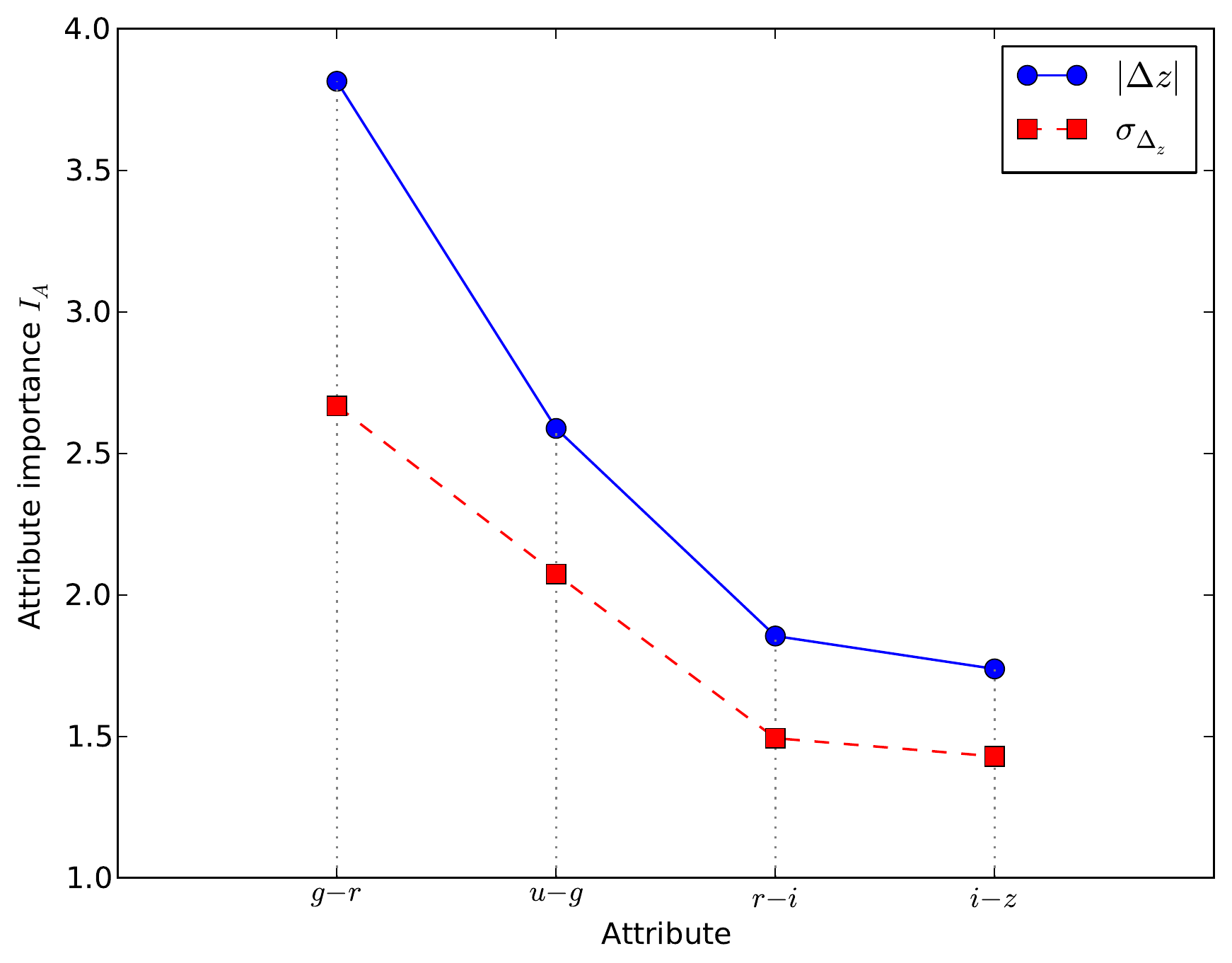}
\includegraphics[width=0.325\textwidth,height=0.3\textwidth]{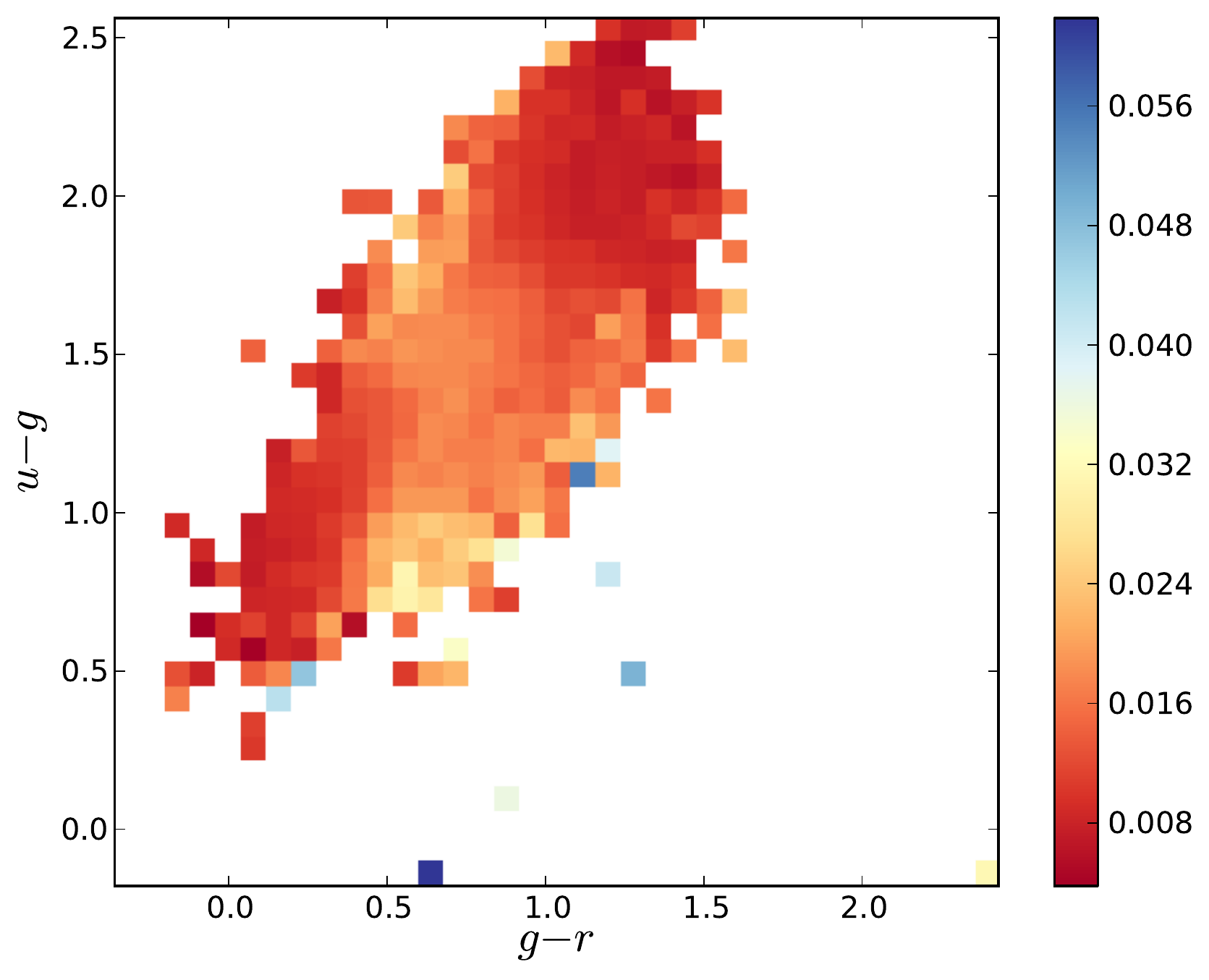}
\caption{(\textit{Left}): The attribute importance factor $I_A$ as a function of redshift for the four attributes (SDSS colors) used in this analysis for the bias only. This factor quantifies how much the metrics decrease as we permute the attributes one at a time. (\textit{Central}): RMS of the relative importance factor as a function of the attributes computed by using the bias (blue) and the scatter( red). (\textit{Right}): A heat map constructed by using the two most important attributes, which indicates areas of parameter space where the \pz prediction is poor. The higher the value  (\ie bluer) in a region, the more training data are needed to increase the accuracy of \pz estimation within that region. These zones might also contain outliers or galaxies with bad photometry.}
\label{fig:vi_sdss}
\end{figure*}

The OOB data can also be used to compute the relative importance of each attribute, which can be done by permuting each of the attributes in the non OOB data when training the tree. The result of this process can be directly compared with the unperturbed case using the OOB data, as shown in Figure~\ref{fig:vi_sdss}. In this figure, the left panel shows the relative importance factor, which is computed by using the absolute value of the OOB bias as a comparison metric, of the four colors used to build the regression trees for the MGS sample. In this plot, a factor of one implies that the attribute acts as a random variable, since a perturbation along that direction produces no changes. Any value greater than one produces a change in the bias, making it larger and therefore less accurate. 

From this figure, we see that the $g-r$ color shows the largest relative importance factor, being close to four, meaning that the absolute bias, on average, changes by this same factor when this color is randomly perturbed. On the other hand, the $i-z$ color is the least, on average, relevant attribute in this context, with a relative importance factor less than 1.5. Due to the limited number of attributes in this test, however, removing this last color actually produces slightly worse results. In the general case when more attributes are present, removing less important variables will improve the results. While this result might seem counter-intuitive, it results naturally from the random nature of the tree construction. Since only $m$ attributes (\eg three) are randomly selected to decide the split dimension, an attribute with overall low importance can be occasionally selected to split a node. By omitting attributes with lower importance, we force the trees to be built from attributes with greater 
information content, thereby improving the accuracy of the prediction.

Another interesting point is that the relative importance for both of the mentioned colors remain consistent, independent of redshift, while the other two colors show variation (\ie $u-g$ and $r-i$ exchange importance ratings more than once), although they are overall consistent with each other. This behavior is mainly due to important spectral features, such as the 4000 \AA\ break, passing between different filters, which TPZ identifies algorithmically, as important indicators of a galaxy's redshift. We see this result from another perspective in the central panel of Figure~\ref{fig:vi_sdss}, which presents the RMS of the relative importance, sorted by their rank, for the four colors computed by using the absolute bias (blue line) and the variance (red line). Both metrics rank the attributes in the same order and either can be used to compute their importance to the data set. Perturbing the attributes produces a stronger effect on the absolute bias than on the scatter, mainly because when perturbing one 
dimension, we lose information and thereby increase the likelihood that a galaxy will end up in a random branch of the tree, especially for an important attribute. This would likely lead to a misclassification, which directly affects the mean absolute bias.

\subsubsection*{Relative Importance}

The importance rank can also be used to better understand  the training data, to check whether it is possible to reduce the dimensionality of the problem, and to identify areas of the mapped parameter space where new training data can be most effectively incorporated. This latter point can be accomplished by identifying the leaf nodes, and the galaxies contained therein, for each tree and computing their accuracy on predicting for the OOB data along with their proximity  matrices. By averaging over these results for all trees, we obtain the desired result. 

For example, by using the two most important attributes previously identified for the SDSS MGS ($g-r$, and $u - g$), we present a heat map in the right panel of Figure~\ref{fig:vi_sdss} that encodes the binned performance of these two attributes, where higher values indicate lower predictive success in that bin. In this plot, we see there are a few bins where performance is markedly lower (blue  and light blue squares), and several areas that are lower than average (the yellow bins). On the other hand, there are two areas where the predictive power is quite high (deep orange-red), which are likely the result of the known color bi-modality of SDSS galaxies~\citep{Strateva2001} where early-type galaxies lie in the upper right part of this plot and late-type galaxies lie in the bottom left part of this plot. The areas in this heat map where the predictive performance is low can be caused by either a lack of training data, by galaxies with color degeneracies, or by galaxies with higher than normal 
magnitude errors. As a result, these areas can be prioritized for follow-up observations to improve the performance of the \pz estimation.

\subsubsection*{Identifying new training data}

Previously, we had stated that the relative importance of the different attributes, graphically shown in the heat map in the right panel of Figure~\ref{fig:vi_sdss}, could be used to optimally identify new training data. We test this assumption by first randomly selecting 1,000 galaxies as our training set, in order to simulate a poor training set, so that we can quantify the effects of both randomly adding new data and selectively adding new data by using the relative importance. We perform this test by first adding 1,000 new galaxies and second by adding 2,000 galaxies and computing the mean normalized bias, defined as $\Delta z'= (z_{\rm spec} - z_{\rm phot})/(1+z_{\rm spec})$, and its standard deviation as we change the size of the training set by using the four color attributes from the SDSS MGS and  and a forest with 100 prediction trees. 

We summarize these test results in Table~\ref{tab:heat_map}. As shown in the table, selecting galaxies from those zones with lower accuracy as indicated by the heat map produces more accurate predictions than adding galaxies randomly. In fact, even adding 1,000 galaxies by using the heat map produces a slightly better performance than adding 2,000  galaxies randomly. These results indicate that it is more important to selectively add galaxies to areas where the prediction is poor than to simply increase the size of the training set. 

We continue this process, by continually adding either 1,000 or 2,000 new galaxies to the training set. As the bottom panel of Figure~\ref{fig:bias_trees} for the SDSS MGS demonstrates, after about 5,000 galaxies (or at half the size of our full training set), the performance metric shows little variation, which is also reflected in the last row of Table~\ref{tab:heat_map} where the metrics for the 15,000 galaxy training set are presented for comparison. This test demonstrates how current and future photometric surveys can optimally construct training sets by either selectively using existing observations, or by  obtaining new spectroscopic observations to improve the \pz estimation.

\begin{table}
\centering
\begin{tabular}{lrr}
\hline
\hline
Number of training galaxies &  $< \Delta z' >$ & $\sigma_{\Delta z'}$  \\ 
\hline
1,000                     & -0.0043    & 0.042 \\
1,000 + 1,000 from random & -0.0037    & 0.038 \\
1,000 + 1,000 from map    & -0.0033    & 0.032 \\
1,000 + 2,000 from random & -0.0034    & 0.036 \\
1,000 + 2,000 from map    & -0.0022    & 0.025 \\
15,000                    & -0.0018    & 0.021 \\
\hline
\end{tabular}
\caption{A comparison of the performance of TPZ for the SDSS MGS when extra data are added  to the training set either randomly or by selectively using ancillary information.}
\label{tab:heat_map}
\end{table}

\subsubsection*{Error distribution}

\begin{figure}
\includegraphics[width=0.48\textwidth]{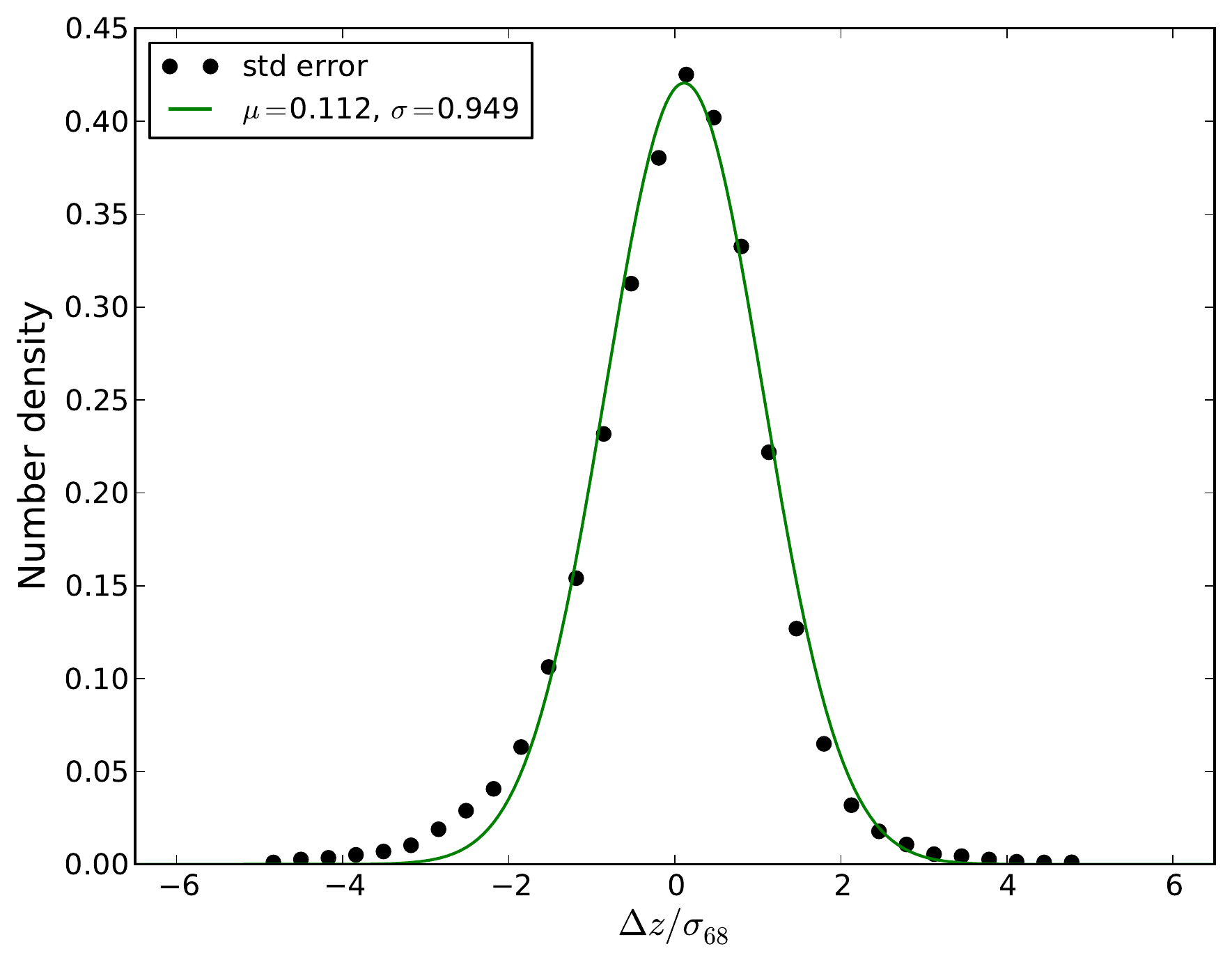}
\caption{The photometric standardized error, $(z_{\rm phot} - z_{\rm spec})/\sigma_{68}$, for the MGS galaxies (black dots) using the mean of each individual PDF and the best fit Gaussian with $\mu = 0.112$ and $\sigma = 0.949$ (solid green curve).} 
\label{fig:std_err}
\end{figure}

After applying TPZ to the SDSS MGS, we can estimate \pz errors directly from the estimated PDF by computing either the mean, the mode, or some other statistic from this distribution. As a demonstration, we calculate the error $\sigma_{68}$ as the region of the \pz PDF centered on the mean that contains 68\% of the cumulative probability. We next calculate the distribution of these standard errors by computing $(z_{\rm phot} - z_{\rm spec})/\sigma_{68}$ for each PDF, which is shown as the black points in Figure~\ref{fig:std_err}. For unbiased standard error estimates, this distribution should be normally distributed with zero mean and unit variance. When we fit our measured points, we obtain a Gaussian with mean equal to 0.112 and a width of 0.949, which is shown by the solid green curve. 

This simple error estimate is quite close to the unbiased expectation, which is as we would expect for any reliable technique. The fit is not a perfect Gaussian due to a slightly extended tail on the left hand side of the distribution. We interpret this as a manifestation of the very narrow PDFs we have obtained and that the SDSS MGS is concentrated at lower redshifts where most \pz techniques suffer from a small tendency to over-predict the photometric redshifts, as shown in the left panel of Figure~\ref{fig:true_plot}. 

\subsubsection*{Size of forest}

When we construct a forest for prediction, one parameter that must be specified is the number of trees that should be constructed. This is important as the more trees in the forest, the higher the computational demands, which slows the training process and construction of \pz PDFs. Thus, we test the performance of TPZ for the SDSS MGS by varying the number of trees built for our forest for a fixed-size training sample. As before, we compute the mean of the absolute bias and its standard deviation, and present how these quantities vary as the number of trees in our forest changes for a fixed training size of 10,000 galaxies. 

These results are presented in the top panel of Figure~\ref{fig:bias_trees}, which shows that our algorithm does become more accurate as the number of trees increases. However, after around 100 trees, the predictive power of the forest shows little variation, indicating that this is a reasonable number of trees for this prediction process. \cite{Breiman2001} demonstrated that, as the number of trees in a random forest increases, any margin function will converge to a limit value. Thus, as expected, we see our generalized error value converging. As a result, this implies that our technique does not over-fit the data as more trees are added in comparison to other machine learning methods.

\subsubsection*{Training size}
\begin{figure}
\includegraphics[width=0.48\textwidth]{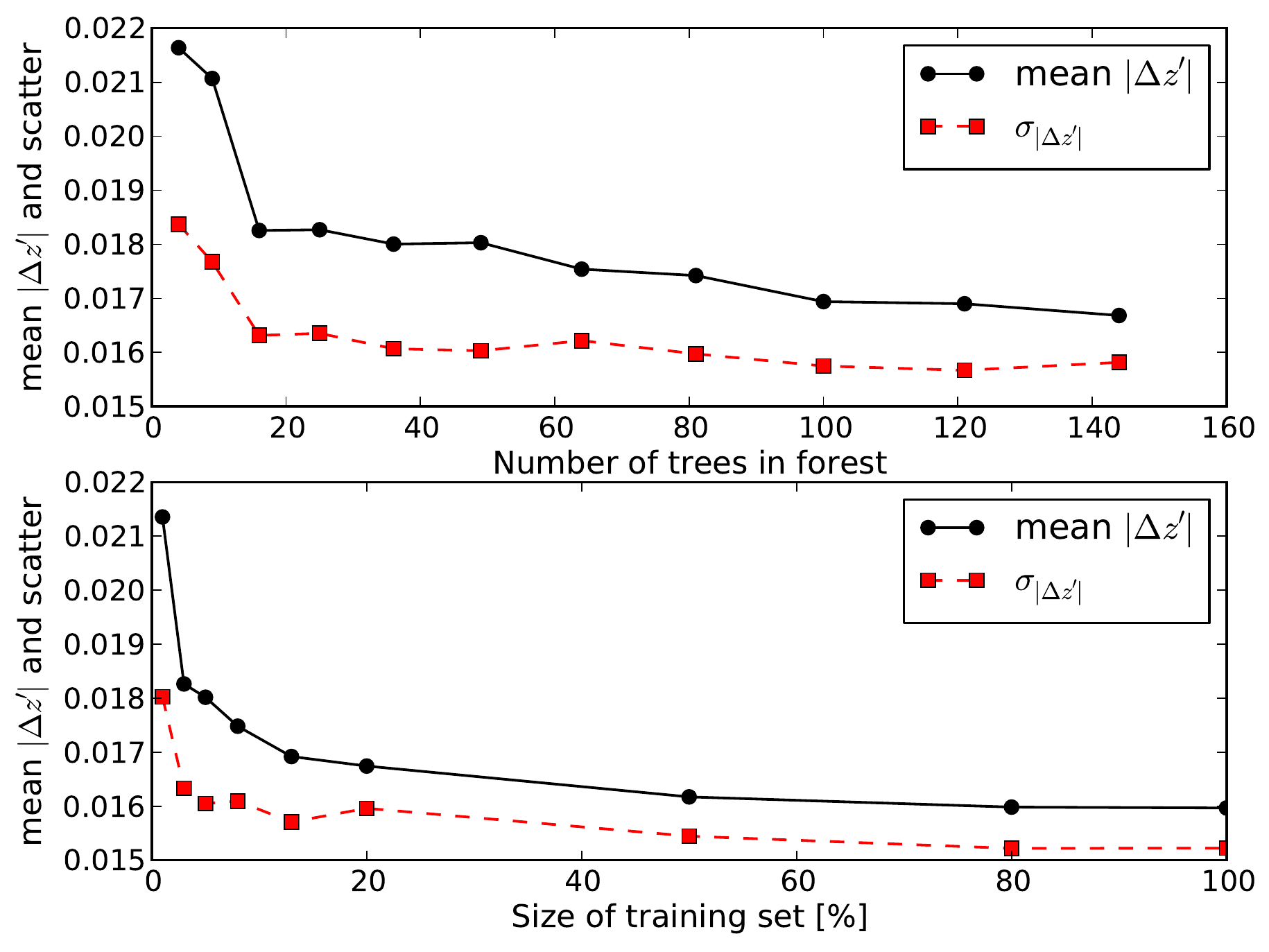}
\caption{The absolute mean normalized  bias defined as $|\Delta z'|= |(z_{\rm spec} - z_{\rm phot})|/(1+z_{\rm spec})$, and its scatter as a function of the number of trees in the forest, keeping the training set fixed (top). The same two values as a function of the size of the training set keeping the number of trees fixed at 100 for galaxies in the SDSS MGS (bottom).} 
\label{fig:bias_trees}
\end{figure}

Once we know the optimal number of trees that must be built for our forest, we next need to know the optimal size of our training set. By using 100 trees (as determined in the previous section), we vary the size of our training set and present the results in the bottom panel of Figure~\ref{fig:bias_trees}. As shown in this figure, the accuracy of TPZ for predicting \pz does not change significantly after using around 70\% of the galaxies. This is an interesting result, that our approach quantifies in an elegant manner, but which will obviously vary between different data sets. Fundamentally, as the training set increases in size, the prediction accuracy also increases until most of the multi-dimensional parameter space has been sampled and little extra information is added by new training galaxies. 

Of course in this test we have not used the relative importance of our parameter attributes, as shown, for example, in the central panel of Figure~\ref{fig:vi_sdss}. By manually selecting additional data, we should be able to reduce the values of these metrics significantly, which is discussed in the next section. But even in our current approach, we expect that some of our test data are not well represented in our training set, which will limit the accuracy of this approach. We see this as an opportunity, however, as we can compute a cross-data proximity matrix by using the trained forest to identify galaxies within the test data that are isolated with few neighbors in the parameter space. Once identified, these galaxies could be treated individually by using, for example, other \pz estimation techniques (see, \eg~Carrasco Kind \& Brunner 2013, in preparation).

\subsection{PHAT1 blind test}

We also tested TPZ on the PHAT1 dataset, described in \S\ref{phat1data}, which is a blind contest where the test spectroscopic redshifts are unknown to the competitors. Therefore, this provides a reliable method to compare the performance of different \pz techniques. In this contest, only a limited quantity of training data are provided; we have approximately 500 galaxies to train our algorithm for the approximately 1,500 galaxies that form the validation sample. These data also have a sparse redshift distribution, extending from $z \approx 0$ to $z \sim 6$. Despite these limitations, we applied TPZ to this training data, submitted our results to the contest, and obtained the resulting performance metrics from the PHAT leader (H. Hildebrandt, private communication). We present our specific results in Table~\ref{tab:phat1}, which can be compared directly with the results shown in Table 5 of the PHAT paper~\citep{Hildebrandt2010}. 

We computed validation results for four different photometric samples: by using all eighteen photometric bands, by omitting the Spitzer photometry and using only fourteen photometric bands, and by creating magnitude limited ($R < 24$) for each of these two galaxy samples. For these validation runs, we use the regression mode to create a forest of 150 trees with $m_* = 4$ (as described in \S\ref{RFs}). In all runs, we made no cuts on the $zConf$ parameter so that we could more directly compare our results to the other competitors. In the end, the TPZ results are among the most accurate \pz predictions, especially when compared to other empirical training codes. Interestingly enough, TPZ even outperforms some template \pz techniques, which are supposedly better suited for this particular challenge due to the dearth of training data and large redshift range covered by the validation sample. These results show that even in less than ideal conditions, 
TPZ provides a robust estimation of photometric redshifts. Note that due to the lack of 
training data and the extended redshift distribution of the validation sample, we did not generate ancillary information for the data by using the OOB approach. 

\begin{table}
\begin{minipage}[]{\columnwidth}
\caption{The TPZ results for the PHAT1 catalogue both with and without the IRAC bands, and for all galaxies and for a magnitude-limited sample with R $<24$. Note that these are the same statistics presented in Table 5 of Hildebrandt et al. (2010) for other \pz estimation techniques.}
\label{tab:phat1}
\centering
\renewcommand{\footnoterule}{}
\begin{tabular}{|l|rrr|}
\hline
\hline
 Run & bias\footnote{bias is defined as: $\Delta z'=\frac{z_{\rm spec}-z_{\rm phot}}{1+z_{\rm spec}}$}& scatter\footnote{RMS of the bias $\Delta z'$} & outlier rate\footnote{Outliers are defined as objects with $|\Delta z'| > 0.15$.} \\
\hline
18-band          & $-0.002$ & $0.055$ & $14.1$ \% \\
14-band          & $-0.007$ & $0.055$ & $12.6$ \% \\
18-band R $<$ 24 & $-0.004$ & $0.055$ & $11.1$ \% \\
14-band R $<$ 24 & $-0.009$ & $0.054$ & $9.6 $ \% \\
\hline
\end{tabular}
\end{minipage}
\end{table}

\subsection*{DEEP2}\label{DEEP2pz}

\begin{figure}
\includegraphics[width=0.46\textwidth]{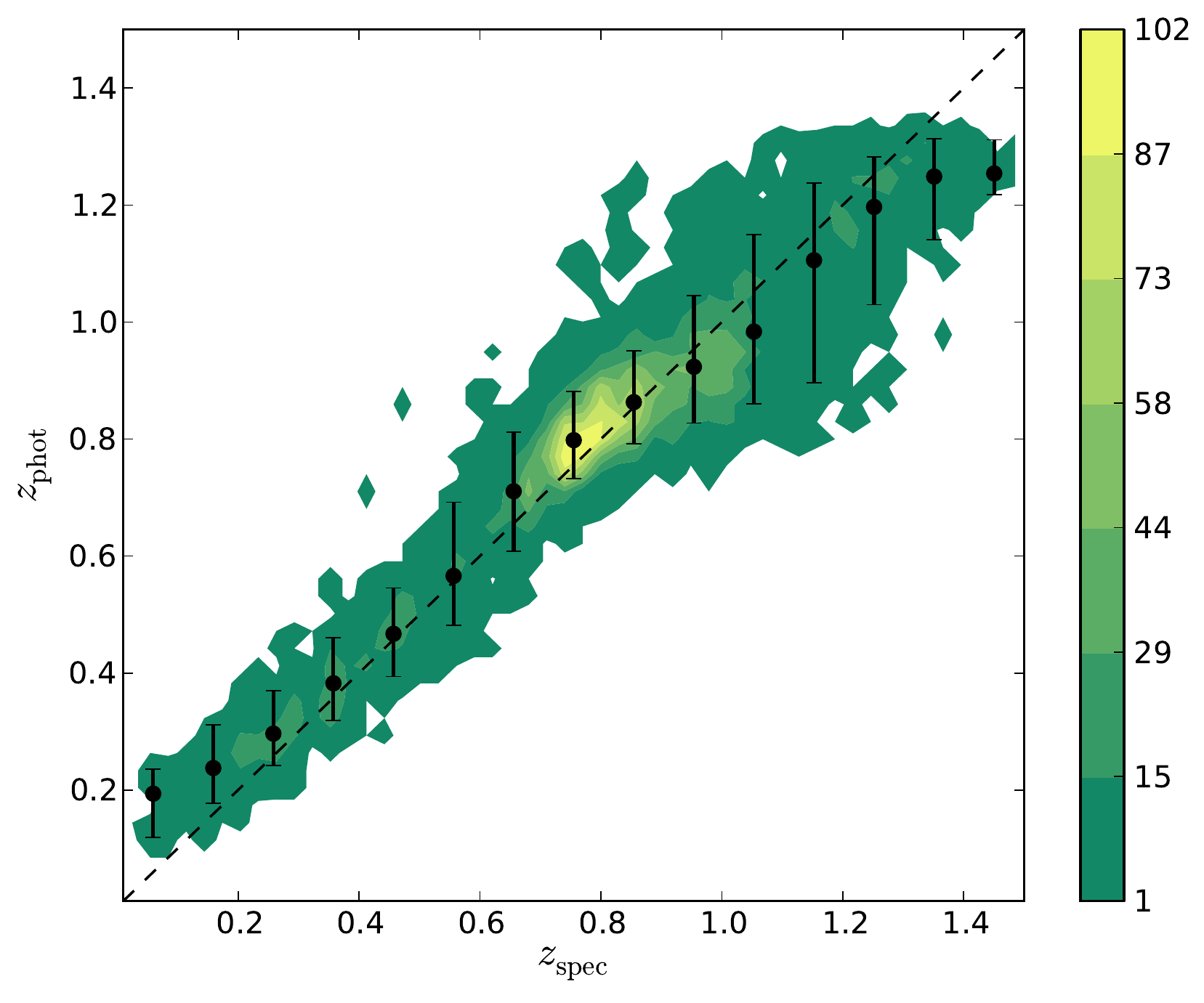}
\caption{The TPZ \pz  with $zConf >$ 0.7 versus the spectroscopic redshifts for 7,856 galaxies selected from the DEEP2 redshift survey. The black dots are the median values of $z_{\rm phot}$ and the errors bars correspond to the tenth and ninetieth percentiles within a given spectroscopic bin of width $\Delta z = 0.1$.}
\label{fig:true_deep}
\end{figure}

We have also tested TPZ by using the DEEP2 redshift survey data, which extends to much higher redshifts than the SDSS MGS. As described in \S\ref{deep2data}, we treat the galaxies with CFHTLS photometry independently from those with SDSS photometry, but in the end we merge the \pz results. We follow a similar analysis to what we used with the SDSS MGS, and after we compute the \pz PDFs, we select only those galaxies with $zConf > 0.7$, which includes about 81\% of the galaxies. We have that the average bias, using  $\Delta z'=(z_{\rm spec}-z_{\rm phot})/(1+z_{\rm spec})$, is -0.007 with $\sigma_{\Delta z'}$ = 0.059 and a outlier rate, defined as $|\Delta z'| > 0.15$ = 2.9\%. We know of no previous \pz analyses of these data (described in~\S\ref{deep2data}) with which to compare these results. The results are presented in Figure~\ref{fig:true_deep}, which compares the \pz computed by using the mean of each individual PDF with the spectroscopic redshift for the 7,856 galaxies. In this figure, we also compute 
the median, shown by the black dots, and the tenth and ninetieth percentiles, shown by the black error bars, within spectroscopic bins of width $\Delta z = 0.1$.

As this figure demonstrates, we see consistent results across all redshifts, and both the isodensity contours and the errors bars indicate that there are few outliers or catastrophic \pzns. However, at both ends of the distribution, we see several bins that show that the \pz results are less accurate and are systematically higher for the first two bins and systematically lower for the last two bins. This effect is often seen with empirical techniques, as the spectroscopic training samples are often less complete at these redshifts, see, \eg the redshift distribution in Figure~\ref{fig:N_z}. Another effect causing this skewness is that estimated photometric redshifts can not be negative, thus our probability distribution can not be symmetrical at the low redshift end. Another possible explanation for the low redshift systematic 
is the effect of galaxy inclination and the induced  extinction on \pz prediction as shown recently by~\cite{Yip2011}.

Likewise, the systematic underestimation at higher redshifts is likely affected by the fact that many of these galaxies are near the limit of the photometry and thus have higher than average magnitude errors. In combination with the lower density of training data, this will reduce the efficacy of our \pz technique. To understand this effect, recall that our trees are built from objects whose photometry is sampled by assuming a normal distribution defined by the magnitude and magnitude error from the bootstrap samples. As the magnitude error increases, the range of possible values to sample increases, thereby producing a sparser sampling for this galaxy within our forest. Since there are few galaxies with redshifts above 1.3 in the training data, the branches on the forest for high-$z$ galaxies are mainly dominated by training galaxies with redshifts closer to 1. As we build the PDF for the high-$z$ galaxies, the PDFs will be positive skewed, and thus the mean value of each PDF will tend to be at lower 
redshift values.

\begin{figure}
\includegraphics[width=0.46\textwidth]{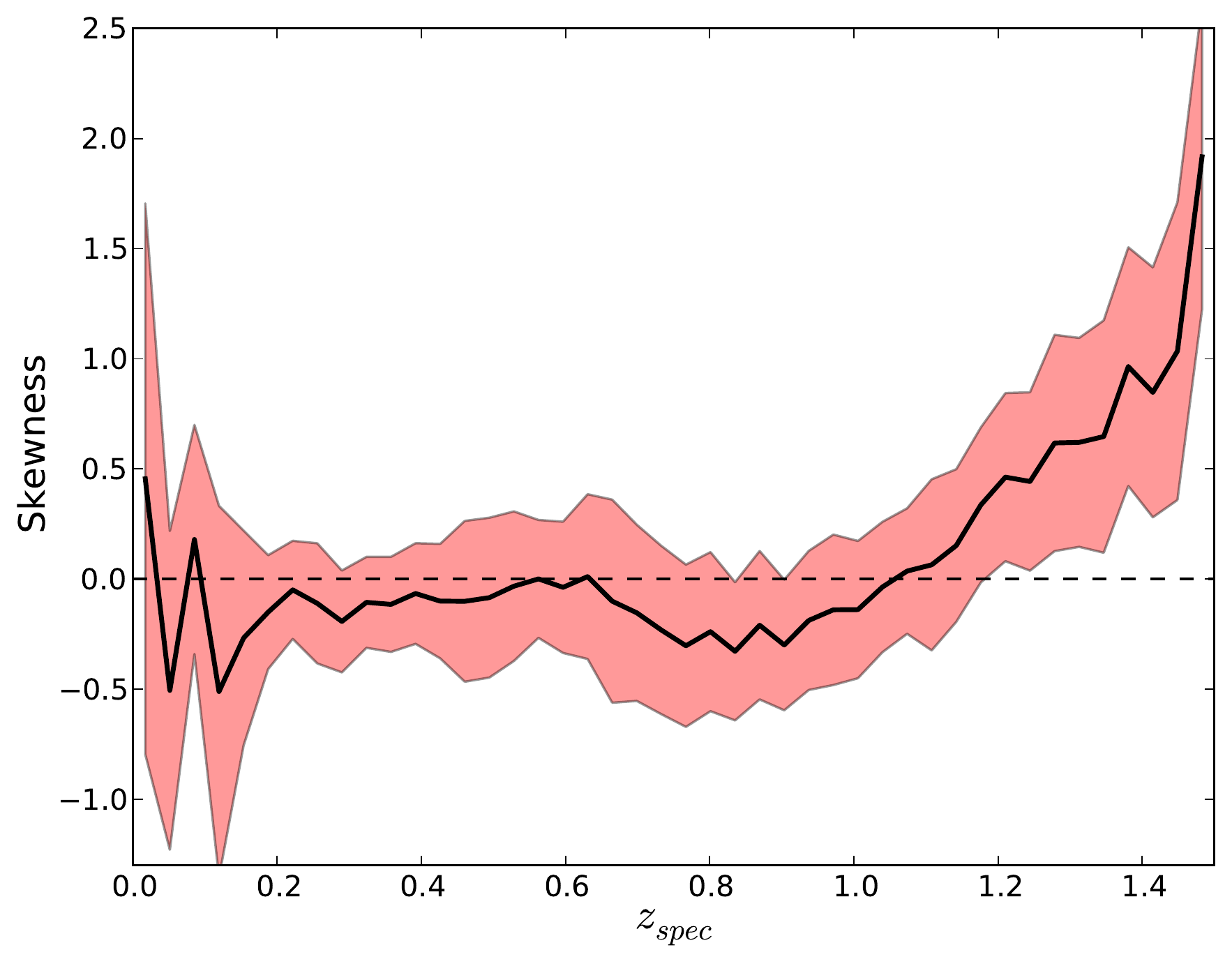}
\caption{The skewness of the \pz PDF as a function of spectroscopic redshift. The solid black line is the mean of the skewness and the pink shaded region corresponds to the one-$\sigma$ interval. Positive skewness indicate a PDF skewed to lower redshifts.}
\label{fig:skew}
\end{figure}

We demonstrate this skewness in Figure~\ref{fig:skew}, which shows the average skewness of the \pz PDFs and the one-sigma error as a function of the spectroscopic redshift. These two quantities are computed as the third standardized moment:
\begin{equation}
 S_k=\int \left(\frac{z-\bar{z}}{\sigma_z}\right)^3 p(z) dz
\end{equation}
with,
\begin{equation}
 \bar{z} = \int z p(z) dz \qquad  \rm and, \qquad \sigma_z=\int (z - \bar{z})^2 p(z) dz
\end{equation}
where the integrals are computed over the redshift domain, and $p(z)$ is the \pz PDF. We can see that for redshifts up to 1.1 the average skewness is very close to zero, showing a small trend to negative values, which will, on average, produce lower values for the mean \pzns. At higher redshifts, however, there is a clear increase in the average skewness, which will tend to produce lower values for the mean of the PDF. It is important to note that even though these PDFs may be (slightly) skewed, they still predict  sufficient probability near the true redshift, information that is overlooked by other methods that use one point predictions. On the other hand, a catastrophic \pz would have a symmetric PDF centered near the wrong redshift, which is not what we observe here.

\subsubsection*{Relative Importance}

By using OOB data, we have computed the metrics from the training data that we compare in Table~\ref{tab:deep2} to the metrics we obtained from the test data after the \pz distributions were computed. The first two rows of this table show the complete results for all attributes for the DEEP2 galaxies. From this we see that there is strong agreement between the OOB and test data results for both the bias and the variance. 
We also computed the relative importance for the eight photometric bands and the $RG$ attribute, which is the estimated R-band radius of an object in 0.207'' pixels (\ie the sigma of the Gaussian fit to the light distribution). 

\begin{figure*}
\includegraphics[width=0.45\textwidth]{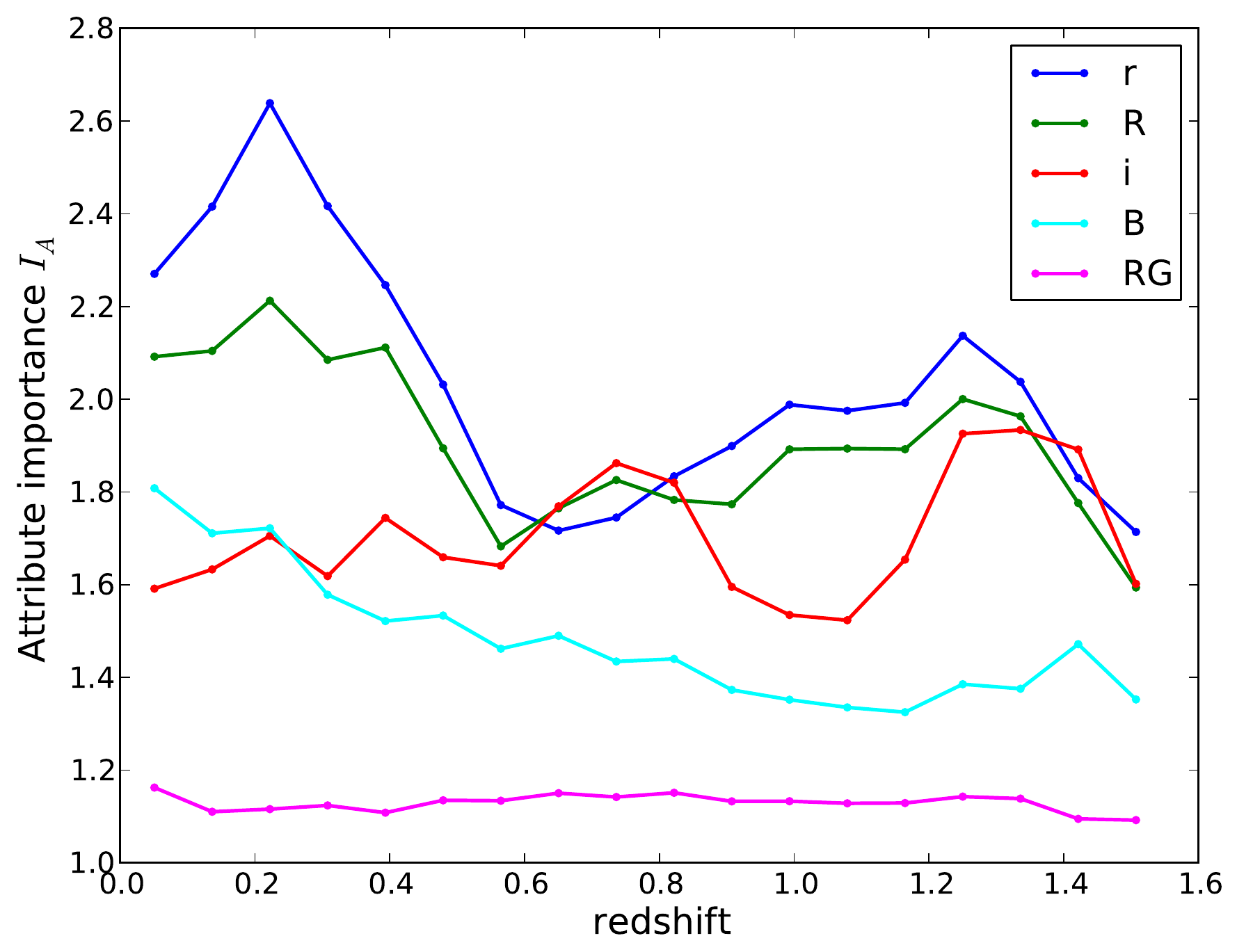}
\includegraphics[width=0.45\textwidth]{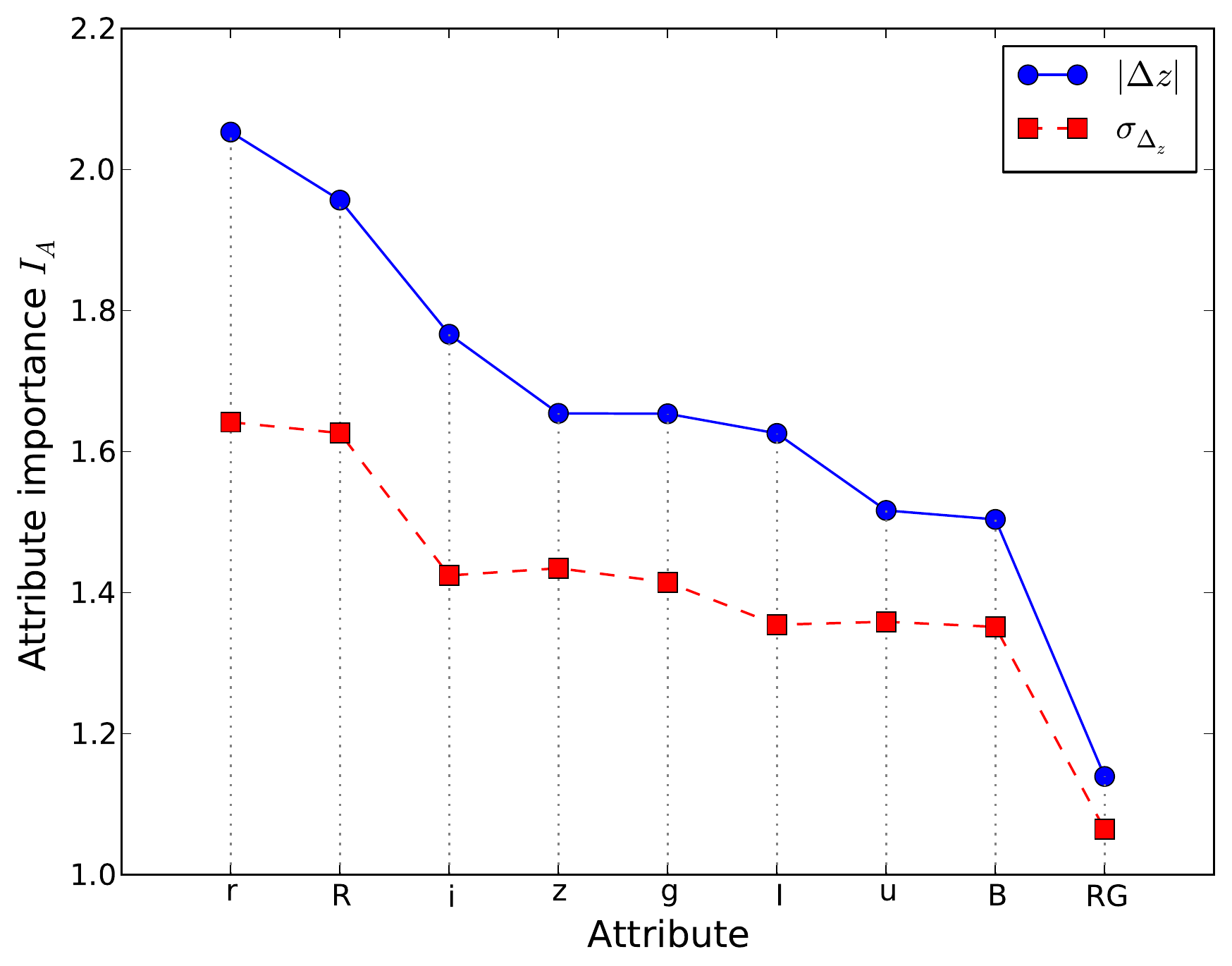}
\caption{(\textit{Left}): The variable importance factor, $I_A$, as a function of redshift for the three most and the two least important attributes (\ie DEEP2 magnitudes) using the bias to quantify this importance. This index specifies how important an attribute is to the calculation of a metric when we permute the attributes one at a time. (\textit{Right}): The RMS of the attribute importance factor as a function of the attributes computed by using the bias (blue) and its scatter (red). Both of these metrics capture the same relative attribute importance.}
\label{fig:vi_deep}
\end{figure*}

In the left panel of Figure~\ref{fig:vi_deep}, we present the attribute importance factor as a function of redshift for the three most and the two least important attributes. From this figure, we see that the R band  and the $r$ band are the most important attributes for making a \pz prediction, similar to the $g-r$ color for the SDSS MGS. This  demonstrates that by a pure statistical analysis, in the optical regime, the R band is the most effective attribute. These attributes show a peak in their importance between redshifts of 0.3 to 0.5. We interpret this increase to the presence of the 4000\AA break being located at these redshifts within these filters. Likewise, the next two most important attributes are the $i$  band and the $z$ band, which are likely important for the same reason, albeit over a slightly higher redshift range.

On the other hand, the least two important attributes are the $B$ band and the $RG$ attribute. As shown in this figure, the $RG$ attribute does not contribute  to the \pz prediction, instead acting like a random variable and thus is likely introducing extra noise into the calculation. We also see no clear evidence that the effect of this attribute changes with redshift. At low redshifts, this attribute could be affected by inclination angle or spectral type, while at higher redshifts, galaxies tend to be fainter and thus have smaller angular sizes. Presumably, these cumulative effects combine to erase any important information this attribute might provide to the \pz calculation.

In the right panel of Figure~\ref{fig:vi_deep}, we present the mean relative importance for each attribute computed from the changes in the mean (blue circles) and the scatter (red squares) when this attributed is permuted, similar to the central panel of Figure~\ref{fig:vi_sdss}. Once again, both metrics agree with the importance ranking. In order to characterize the attributes and their computed ranking of importance, we have made the following tests. First, we removed the two least important attributes, after which we remove the two most important attributes, reincorporating the previously removed attributes. We repeat this process, but now we remove alternately  the four least and most important attributes. In each case, we test whether TPZ is able to correctly recognize the attribute importances. These results are summarized in Table~\ref{tab:deep2}, where we use the absolute mean value of $\Delta z'$ and its dispersion.

\begin{table}
\begin{minipage}[t]{\columnwidth}
\caption{A comparison in the accuracy of \pz predication by using different attribute combinations from the DEEP2 data for all test galaxies. The first row are the metrics for TPZ using only OOB data, which are comparable to the values obtained from the full test data, shown in the second row. The remaining rows provide these metrics for training data that have had the indicated number of attributes removed from the calculation.}
\label{tab:deep2}
\centering
\renewcommand{\footnoterule}{}
\begin{tabular}{l|lrr|}
\hline
\hline
Attribute Selection & $< |\Delta z'|>$ & $\sigma_{\Delta z}$ \\
\hline
All attributes (OOB metrics)  & 0.052 & 0.053 \\
All attributes & 0.047 & 0.049 \\
Remove 2 least important & 0.044 & 0.046  \\
Remove 2 most important & 0.061 & 0.068 \\
Remove 4 least important & 0.044 & 0.048 \\
Remove 4 most important & 0.070 & 0.084 \\
\hline
\end{tabular}
\end{minipage}
\end{table}

As is not surprising, we see that removing the two least important attributes does, in effect, improve the precision of TPZ while also making the code run faster since we have fewer dimensions to check when splitting nodes within the tree, less data to keep in memory when building the tree thus improving cache access, and random realizations from the input parameters will be faster since there are fewer dimensions to sample. Yet, removing four attributes shows a slight decrease in the overall performance, in this case we have removed too much information. While this decrease might seem rather small, since we are randomly selecting attributes when splitting nodes within the trees, by removing four we have increased the scatter since we are losing information. On the other hand, removing the most important attributes significantly affects the results, regardless of how many attributes we remove. As we would expect, the reason is clear. Since these attributes have the most information needed to subdivide the 
multidimensional parameter space in order to produce accurate  \pzns, removing them negatively impacts the performance of TPZ. 

\begin{figure}
\includegraphics[width=0.46\textwidth]{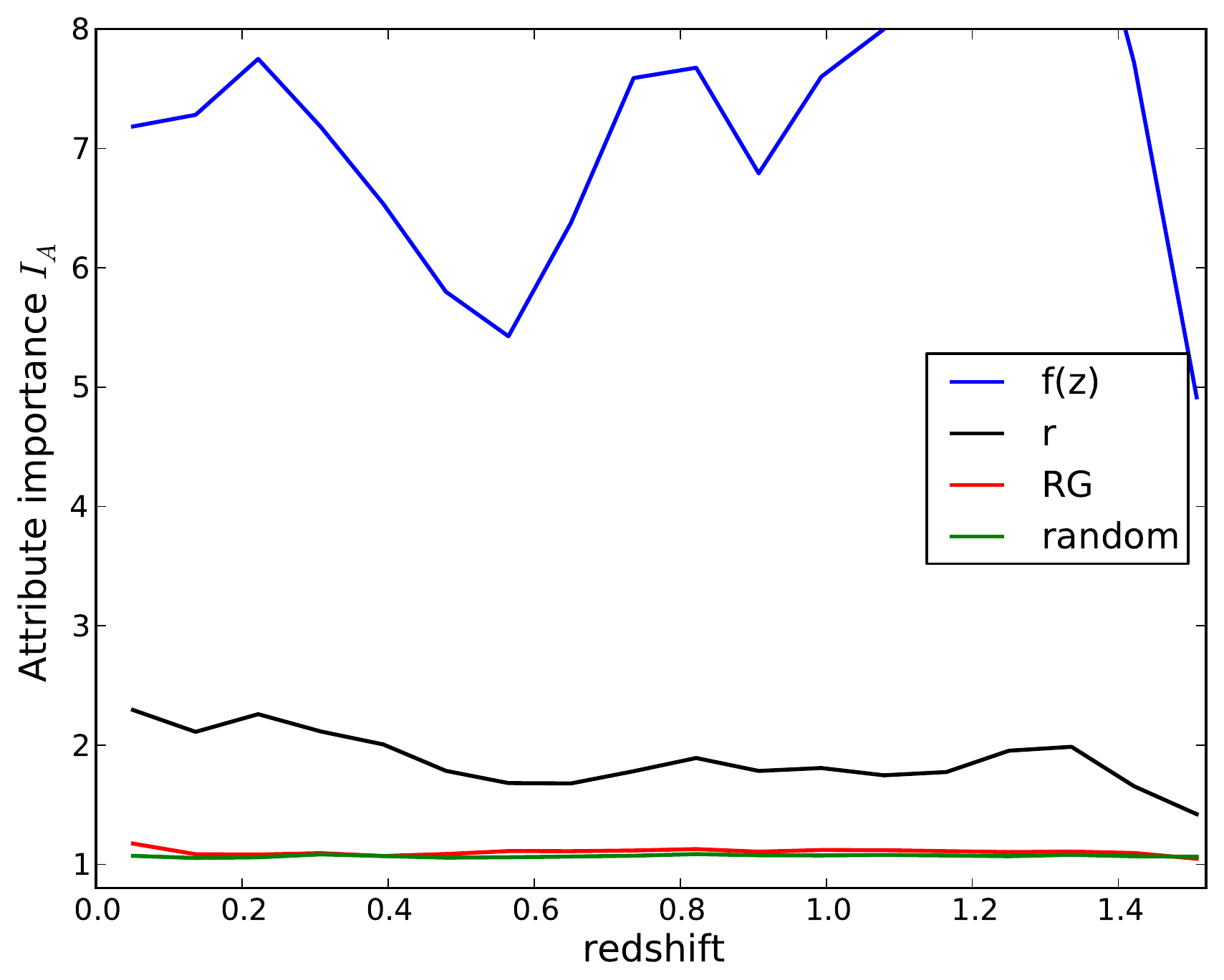}
\caption{The variable importance factor, $I_A$, as a function of redshift for the most and least important attribute using the bias to quantify the importance ranking. As a control test, we added  two artificial variables: an attribute that is a function of the spectroscopic redshift, and a uniformly distributed random attribute. TPZ is able to recognize these two extra attributes and rank them accordingly, as shown by the figure legend.}
\label{fig:added}
\end{figure}

In a further control test we added two extra artificial variables to the data set, one of which is strongly dependent on the source redshift, i.e. a function of redshift, while the second one is a uniformly distributed random variable. After computing their importance rankings, we can see from Figure~\ref{fig:added} that TPZ recognized these two extra attributes and put them on the extreme limits of the importance ranking. The most important value ranks at about eight, the random variable ranks at one as expected, while the $r$ magnitude is  ranked with a value close to two. We notice that the variable $RG$ is very close to one, and therefore to a random variable. As discussed above, we can safely remove this variable from our calculation as it does not provide any useful information. The legend on the plot indicates also the descending order in importance, in concordance with Figure~\ref{fig:vi_deep}.

\subsubsection*{Missing data}

One interesting capability of TPZ is that it can be used to replace attributes in data that are either missing attributes or have attributes with large uncertainties. As discussed in \S\ref{AI}, the replacement values can be computed from the proximity matrix, and we can apply this technique to data either in the training sample or in the application sample. In the former case, missing attributes would be replaced in order to maximize the size of the training set. The alternative would be to simply cull data with missing attributes from the training sample, which would decrease the robustness of our predictive power. In the latter case, missing attributes would be replaced in order to estimate a photometric redshift for a galaxy based on the incomplete but available information. In most cases, this will still result in a reliable prediction, without discarding any data, thereby increasing the overall statistical power of our approach. 

To demonstrate this capability, we selected training and testing data sets that initially were complete and had relatively small errors (i.e., magnitude errors $<$ 1 magnitude). We first randomly replaced 50\% of the magnitudes in the training data with a bad value (e.g., 99), thus some galaxies in this sample have multiple bad attributes. From this new data set, we apply TPZ to generate a second training sample where the bad attributes have been replaced, using only six iterations (i.e, the replaced sample), and we also generate a third training sample where we simply remove any galaxies with missing or bad attributes (i.e., the cut sample). Likewise, we also generated a test sample with 50\% of the attributes replaced by the bad value (i.e., the bad test sample) 

We estimate \pzns s for the clean test sample by using all three training samples: the original, clean sample (i.e., the control), the replaced attribute sample, and the cut sample. Likewise, we use the clean training sample to replace missing attributes and estimate \pzns s for the bad test sample. We present the results of these tests in Table~\ref{tab:missing}, where we compare the \pz estimation for the clean sample with the replaced and cut samples. For this comparison, we use $\Delta z_{pp} = z_{\rm phot,clean} - z_{\rm phot,other}$ and $\Delta mag = mag_{\rm clean} - mag_{\rm other}$, along with their variances, where \textit{other} can either be the replaced or cut samples. As shown in this Table, the replaced value sample produces, on average, superior \pzns s than the cut sample. Likewise, we have estimated robust \pzns s for the bad test sample, which significantly increases the size of our resulting test data. Dealing with missing attributes is important , especially when a spectroscopic training 
sample is limited or 
when cross-matching between incomplete catalogs is carried out in order to develop a more complete catalog for \pz estimation.

\begin{table}
\begin{minipage}[t]{\columnwidth}
\caption{\PZ estimation metrics to demonstrate the robustness of our missing attribute technique. The first two rows show the average bias and its variance between the estimated \pz, and replaced magnitude when either removing or  recovering bad data in comparison with \pzns s predicted using the original, clean sample. The last row shows the the same metrics calculated by using the clean test sample, but for data missing in the test sample as compared to the clean original sample.}
\label{tab:missing}
\centering
\renewcommand{\footnoterule}{}
\begin{tabular}{l|cccc|}
\hline
\hline
Recovered train & $< \Delta z_{pp}>$\footnote{in units of $10^{-3}$} & $\sigma^2_{\Delta z_{pp}}$\footnote{in units of $10^{-3}$}  & $\Delta mag$ & $\sigma^2_{\Delta mag}$\\
\hline
with removed data   & -1.27 & 3.5 & -- & --\\
with recovered data & 0.40 & 1.6 & 0.021&0.094\\
\hline
\hline
Recovered test & $< \Delta z_{pp}>$ & $\sigma^2_{\Delta z_{pp}}$  & $\Delta mag$ & $\sigma^2_{\Delta mag}$\\
\hline
with recovered data &0.72 &4.5&0.033&0.12\\
\hline
\end{tabular}
\end{minipage}
\end{table}

\begin{figure*}
\includegraphics[width=0.45\textwidth]{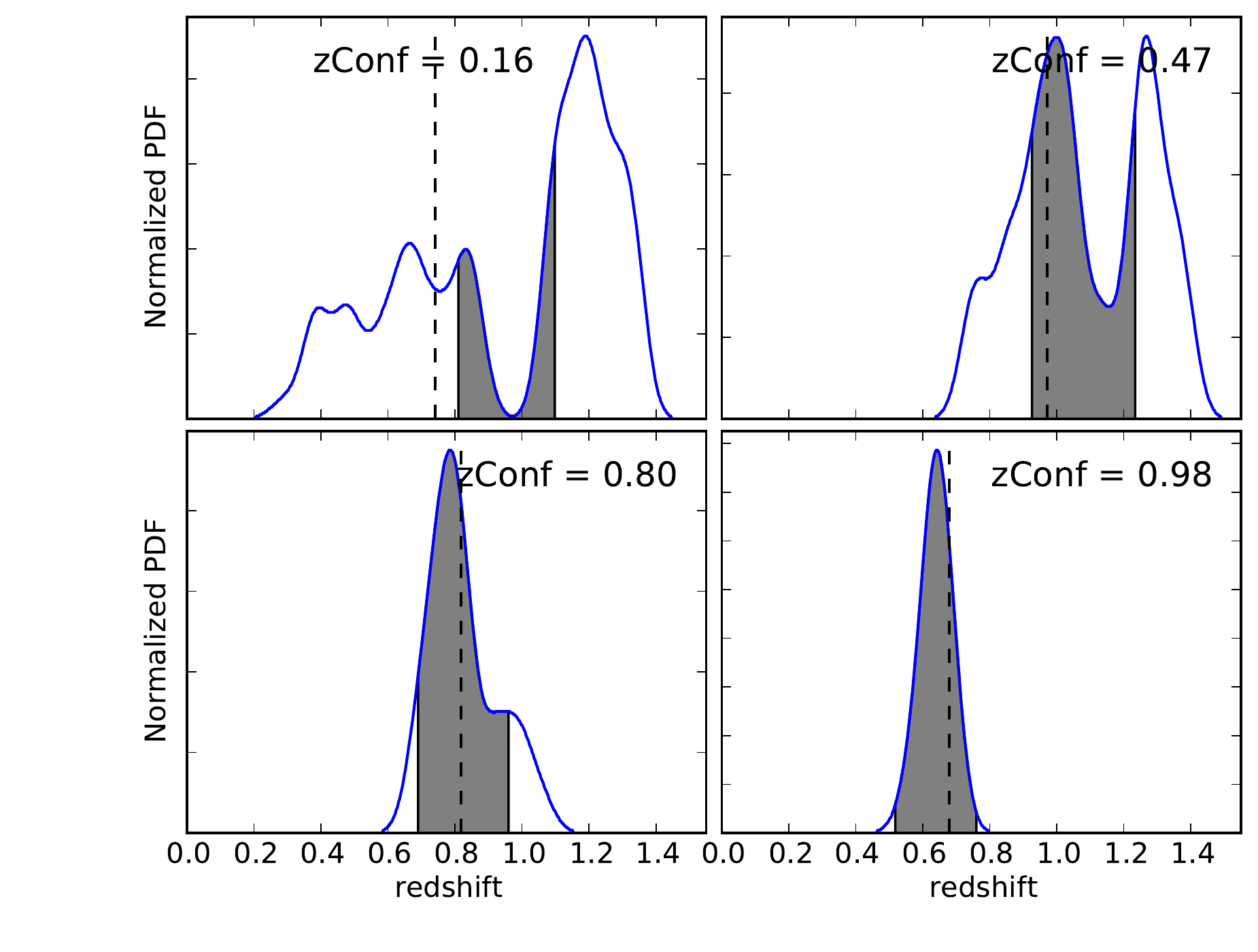}
\includegraphics[width=0.45\textwidth]{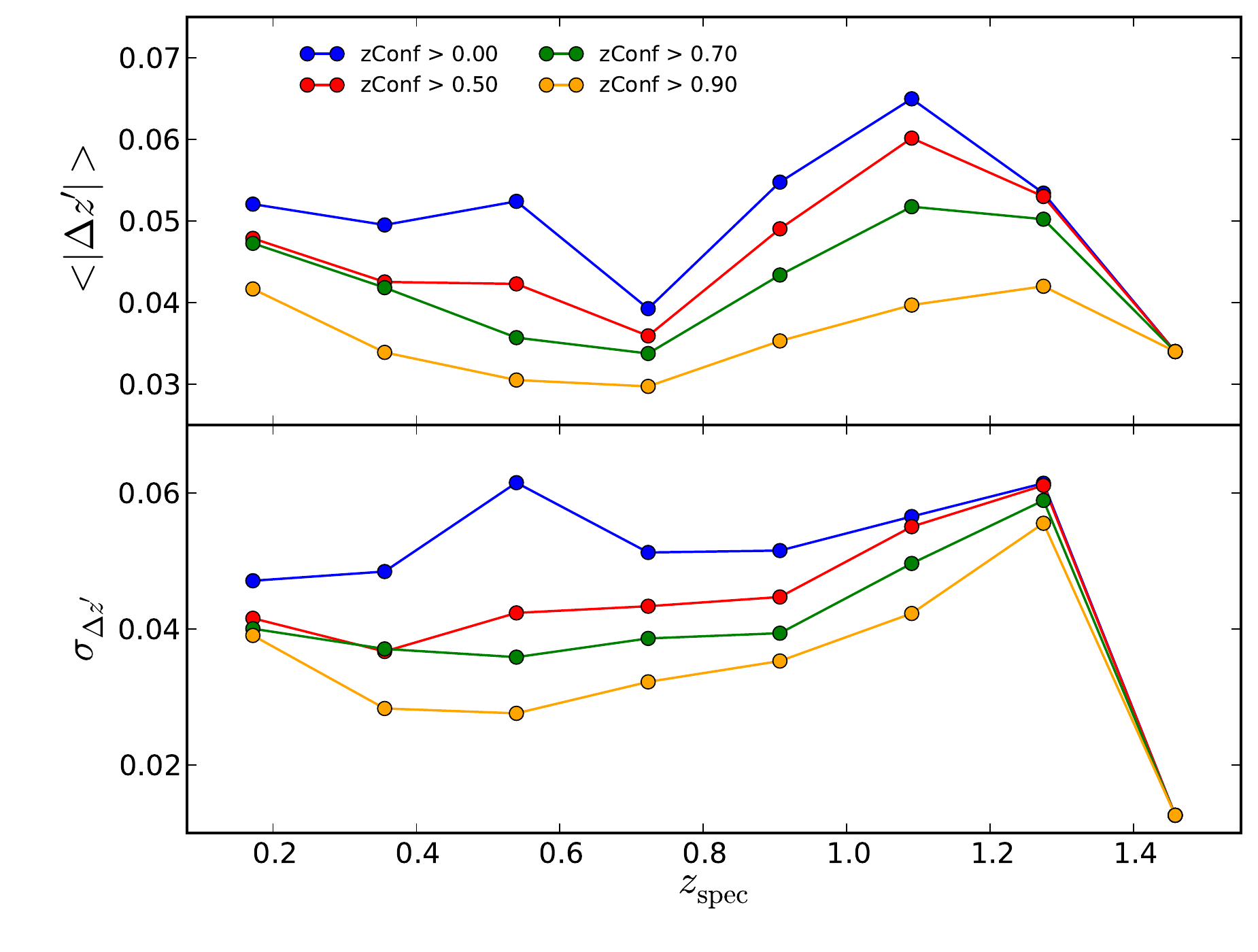}
\caption{(\textit{Left}): Same as Figure~\ref{fig:pdf_ex} but for four example galaxies taken from DEEP2. The vertical dashed line indicates the spectroscopic redshift and the gray area the $zConf$ value. (\textit{Right}): The absolute normalized bias and the scatter for galaxy samples defined by different $zConf$ cuts by using the mean of the \pz PDF as our estimate.}
\label{fig:deep_zconf}
\end{figure*}

\subsubsection*{Photo-$z$ PDFs and $zConf$}

As discussed in  \S\ref{SDSS-MGS}, the $zConf$ parameter can be used to identify galaxies with narrow, concentrated \pz PDFs, which ideally will result in galaxy samples that have the most accurate \pz estimates. The $zConf$ parameter is demonstrated for DEEP2 galaxies in the left panel of Figure~\ref{fig:deep_zconf}, which shows four representative \pz PDFs selected with different values of $zConf$ as measured about the mean of each PDF. Both this figure and Figure~\ref{fig:pdf_ex}, which presents four \pz PDFs by using SDSS data, highlight the fact that wide and sparse distributions have low $zConf$ values while narrower PDFs have higher $zConf$ values. 

The goal of a parameter like $zConf$ is to algorithmically identify galaxies that have, on average, the most accurate \pz estimates. To test this hypothesis, we used all available DEEP2 training data to bud our prediction trees and estimate \pzns s for the DEEP2 test sample. From this sample, we applied three $zConf$ cuts: 0.5, 0.7, and 0.9, and calculated the bias and scatter as a function of redshift for the three resulting galaxy samples. We compare these results to the bias and scatter when no $zConf$ cut is applied in the right hand panel of Figure~\ref{fig:deep_zconf}. As shown in this figure, both the mean absolute bias and the scatter are reduced as $zConf$ is increased, independent of redshift. 

A simple, intuitive approach to select galaxies by their $zConf$ would be 0.5 as this selects galaxies that have a 50\% probability that their \pz redshift estimate lies within the limits imposed by $\pm \sigma_{TPZ}(1+z_{\rm phot})$. Furthermore, higher values would provide more accurate results at the expense of reduced statistical power (i.e., a smaller, final catalog). In Figure~\ref{fig:deep_zconf}, for example, cuts on $zConf$ at 0.5, 0.7 and 0.9 keep 90\%, 76\% and 38\% of the galaxies from the original catalog. Alternatively, given the OOB data predictive results, a required accuracy or number density can be used to identify a suitable value of $zConf$.

\subsubsection*{N(z): An application of \pz PDFs}

Most of the results we have presented within this paper have been based on the estimation of a single metric computed from the \pz PDF, for example the mean or mode. Obviously, using a single value to represent the PDF wastes significant information, but since many \pz applications mimic spectroscopic redshift applications, new approaches must be developed to capitalize on the full information content of a \pz PDF. As a result, we present a simple, yet very important application that uses the full \pz PDFs---estimating the galaxy redshift distribution, N(z). This function is a fundamental measurement and is very important to a number of cosmological applications including weak lensing tomography~\citep[\eg][]{Mandelbaum2008,Jee2012} and projecting three-dimensional theoretical power spectra to angular clustering measurements~\citep{Hayes2012}.

\begin{figure}
\includegraphics[width=0.47\textwidth]{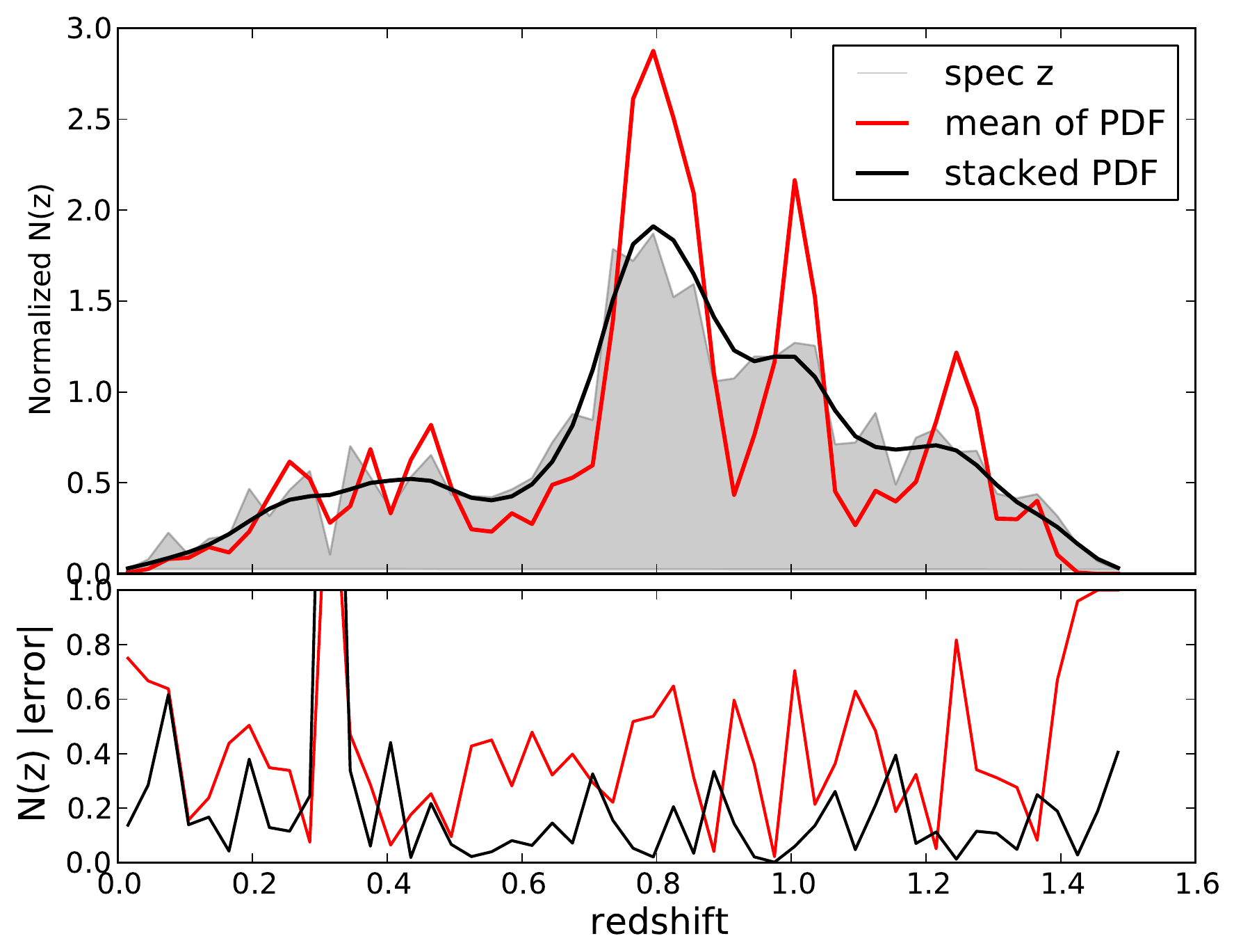}
\caption{(\textit{Top}): The redshift distribution for the all DEEP2 spectroscopic sample of galaxies (shaded gray histogram), computed from the mean value of individual \pz PDFs (red curve), and computed by stacking individual \pz PDFs (black curve). (\textit{Bottom}): The residual absolute error between the spectroscopic redshift distribution and the two \pz redshift distributions shown using the same color scheme.}
\label{fig:N_z}
\end{figure}

We compute the normalized galaxy redshift distribution, $N(z)$, for all the galaxies in  DEEP2 sample (\ie no DEEP2 redshift confidence selection cut was applied), shown in Figure~\ref{fig:N_z} as the shaded gray area. As demonstrated by this figure, in this spectroscopic survey, most galaxies were selected to have redshifts between 0.6 and 1.2. Next, we compute the binned photometric redshift distribution by using the mean value from each \pz PDF, shown by the red curve. While this curve does trace the gross features of the underlying spectroscopic redshift distribution, it fails to capture the full detail and can be significantly different at certain redshifts, including at the mode. For comparison, we show in black the \pz PDF redshift distribution that we obtain by simply stacking the individual PDFs together. With this simple approach, we obtain a more accurate representation of the true sample redshift distribution. Here we have used all the galaxies, without selecting galaxies by their confidence 
level. This demonstrates that all individual PDFs computed with TPZ carry important information about the underlying distribution.

These differences are more clearly exposed in the bottom panel of Figure~\ref{fig:N_z}, where we show the absolute fractional error, $(N_{phot}-N_{spec})/N_{spec}$, as a function of redshift, using the same color scheme as before. From this figure, we see that the stacked PDF has a smaller error for almost all redshifts. In addition, the \pz PDF redshift distribution is considerably smoother and looks more like a fit to the spectroscopic sample, which is another benefit of using the full \pz PDF. For this particular demonstration,  the \pz PDF presented used a bin size of 0.002, while the spectroscopic and photometric redshift distributions used a bin size of 0.03. Of course, we can generate smoother distributions for either the spectroscopic or \pz mean value redshift distributions by reducing the bin size, however, the trade off is that we run the risk of increasing the shot noise in the resulting distribution.

\section{Conclusions}

In this work we have presented and publicly release\footnote{http://lcdm.astro.illinois.edu/research/TPZ.html} TPZ, a new, parallel, machine learning  \pz Python code that computes \pz PDFs and  also provides ancillary information about the photometric data. TPZ is based on the construction of prediction trees and consequently a random forest. Overall, TPZ is a three step algorithm that first preprocesses the data, completes galaxies with missing photometric values in an efficient manner, and also incorporates measurements errors. A \pz PDF can be generated from the prediction trees in one of two modes: classification or regression. Both modes produces similar accuracies, but the regression mode is preferred when either the training data are either poorly sampled or not uniformly distributed.  On the other hand, the classification mode provides a detailed synopsis of the redshift distribution that can be used to construct priors for use with other \pz techniques.

We demonstrated the efficacy of the TPZ algorithm and its implementation by applying this new code to three different data sets: the SDSS main galaxy sample, the PHAT1 blind challenge, and the DEEP2 survey. With the SDSS MGS, we demonstrated that using confidence levels is important as they improve the overall accuracy of our \pz sample by selecting those galaxies with narrow PDFs. This technique is unique in the sense that it does not need a separate validation test, yet provides ancillary information by using OOB data. We have shown that with these data, we obtain unbiased estimates of both the bias and the dispersion, which are very similar to the same values obtained from the test data for both the SDSS MGS and DEEP2. Obviously, this result is extremely important when working with data that have unknown redshifts.

TPZ not only provides these prior metrics, but it also provides a ranking of the relative importance of the different photometric attributes that are used by the code. This completely statistical process recovers what is naturally expected from physical consideration of these different attributes. With this importance ranking, we can construct a heat map of the different locations in parameter space that produce poor \pz estimations. Furthermore, we demonstrated that by adding new, manually selected data we can produce more accurate \pz predication than by simply adding new galaxies randomly. This implies that we can optimally identify new training data for current and future photometric surveys, such as DES or LSST, in order to improve their \pz predictions.

The attribute importance can also be used to remove those attributes that are least important, thereby improving the computational speed. In addition, we demonstrated that the performance metrics converge as the number of trees increases in the forest, providing a further method to reduce the computational time since we have a direct measure of the minimum forest size. Likewise, we also demonstrated by using the SDSS MGS that these same metrics also converge with the number of training galaxies for a fixed forest size. Thus, except for adding in manually selected training data to improve areas with poor \pz prediction, we have an explicit limit for the number of training galaxies needed. Finally, with this technique we found that the error distribution was characterized by a Gaussian distribution with a mean very close to zero and variance very close to one, indicating that the source of errors is relatively unbiased.

We ran our code on the PHAT1 blind challenge with excellent results; even with limited training data we were able to compute accurate \pz's that were comparable if not better to other empirical techniques as well as to some SED fitting techniques. By using the DEEP2 redshift data, we tested TPZ over a large redshift range, obtaining very accurate results. In particular, we were able to identify the important attributes, which in this case was the R band magnitude followed by the I band magnitude, and the least important attributes, which in this case was the $RG$ attribute and the $B$ band magnitude. Despite these impressive results, we still have a slight systematically biased \pz at very low and very high redshifts, which we primarily believe is caused by the low number of training data at these redshifts and also the fact that \pz estimates can not be negative. We also see a positive skewness in the \pz PDFs at high redshifts. We believe this result is due to the fact that these galaxies tend to be 
fainter and have larger magnitude errors. These larger magnitude errors produce a sparser forest at higher redshifts, which is manifested by having a lower \pz PDF mean value at these same redshifts.

We have also demonstrated how the $zConf$ parameter can be used to select galaxy samples that have improved \pz estimates with minimal outliers. A target value for this useful parameter can be set to a desired \pz precision either by calculating the value expected by using OOB data or as required by a specific cosmological requirement. Likewise, we have demonstrated how TPZ can efficiently handle missing data within a catalog. By artificially generating bad or missing parameter values within both the training and the testing data sets, we were not only able to robustly recover the missing parameters but more importantly new \pz estimates that are consistent with the \pz estimates from the original, full data set. Therefore, this technique increases the power of \pz estimation by recovering missing data from the training catalog as well as the power of our resulting sample statistics by recovering missing data from the application data set.

Finally, by calculating the normalized distribution of galaxies as a function of redshift, we were able to demonstrate the advantages of using a full \pz PDF as opposed to using one single estimator of the PDF or any other point metric. Specifically, by simply stacking each individual PDF, we recover the underlying galaxy redshift distribution to a much higher precision than by simplifying using the mean of each individual \pz PDF.

In conclusion, we note that since TPZ is an empirical algorithm, it is inherent dependent on the quality of its training data. Thus, as is the case with all empirical algorithms, TPZ is limited by the available spectroscopic training data. Furthermore, the application of TPZ to regions of parameter space beyond the limits of the training data (i.e., extrapolation) will be less reliable. We do note, however, the TPZ does provide ancillary information that can be investigated to better understand the limitations imposed by the training set, to identify the optimal locations within the application data space where new training data will be most useful, and to quantify the possible errors associated with the extrapolation of this technique. Another approach to improve the efficacy of \pz estimation beyond the limits of a spectroscopic training sample would be to include the predictions from different, non-empirical techniques into a meta-classifier. We will explore this approach in a future work on this topic.

\section*{Acknowledgements}

The authors thank the referee for a careful reading of the manuscript. The authors also gratefully acknowledge the use of the parallel computing resource provided by the Computational Science and Engineering Program at the University of Illinois.  The CSE computing resource, provided as part of the Taub cluster, is devoted to high performance computing in engineering and science. This work also used resources from the Extreme Science and Engineering Discovery Environment (XSEDE), which is supported by National Science Foundation grant number OCI-1053575.

MCK has been supported by the Computational Science and Engineering (CSE) fellowship at the University of Illinois at Urbana-Champaign. RJB has been supported in part by the Institute for Advanced Computing Applications and Technologies faculty fellowship at the University of Illinois.

Funding for the DEEP2 Galaxy Redshift Survey has been
provided by NSF grants AST-95-09298, AST-0071048, AST-0507428, and AST-0507483 as well as NASA LTSA grant NNG04GC89G.

The SDSS is managed by the Astrophysical Research
Consortium for the Participating Institutions. The Participating Institutions are the American Museum of Natural History, Astrophysical Institute Potsdam, University of
Basel, University of Cambridge, Case Western Reserve University, University of Chicago, Drexel University, Fermilab,
the Institute for Advanced Study, the Japan Participation
Group, Johns Hopkins University, the Joint Institute for
Nuclear Astrophysics, the Kavli Institute for Particle Astrophysics and Cosmology, the Korean Scientist Group, the
Chinese Academy of Sciences (LAMOST), Los Alamos National Laboratory, the Max-Planck-Institute for Astronomy
(MPIA), the Max-Planck-Institute for Astrophysics (MPA),
New Mexico State University, Ohio State University, University of Pittsburgh, University of Portsmouth, Princeton
University, the United States Naval Observatory, and the
University of Washington

\bibliographystyle{mn2e}
\bibliography{refDTpaper}

\bsp
\label{lastpage}
\end{document}